%% file: main.tex
\documentclass[sigconf, screen, submit]{acmart}
\usepackage[dvipsnames]{xcolor}
\AtBeginDocument{%
  }

\setcopyright{none} 
\copyrightyear{2025}
\acmYear{2025}
\acmDOI{XXXXXXX.XXXXXXX}

\acmConference[arXiv 2025]{}{}{}

\acmISBN{978-X-XXXX-XXXX-X/XX/XX}

\begin{document}

\title{CompAir: Synergizing Complementary PIMs and In-Transit NoC Computation for Efficient LLM Acceleration}


\author{Hongyi Li}
\affiliation{%
  \institution{Center for Brain-Inspired Computing Research, Tsinghua University}
  \country{Beijing, China}
}

\author{Songchen Ma}
\authornotemark[1]
\affiliation{%
  \institution{The Hong Kong University of Science and Technology}
  \country{Hong Kong, China}
}

\author{Huanyu Qu}
\affiliation{%
  \institution{University of Macau, Guangdong Institute of Intelligence Science and Technology}
  \country{Macau, China}
}

\author{Weihao Zhang}
\affiliation{%
  \institution{Center for Brain-Inspired Computing Research, Tsinghua University}
  \country{Beijing, China}
}

\author{Jia Chen}
\affiliation{%
  \institution{The Hong Kong University of Science and Technology}
  \country{Hong Kong, China}
}

\author{Junfeng Lin}
\affiliation{%
  \institution{Center for Brain-Inspired Computing Research, Tsinghua University}
  \country{Beijing, China}
}

\author{Fengbin Tu}
\affiliation{%
  \institution{The Hong Kong University of Science and Technology}
  \country{Hong Kong, China}
}

\author{Rong Zhao}
\authornotemark[1]
\affiliation{%
  \institution{Center for Brain-Inspired Computing Research, Tsinghua University}
  \country{Beijing, China}
}

\begin{abstract}

The rapid advancement of Large Language Models (LLMs) has revolutionized various aspects of human life, yet their immense computational and energy demands pose significant challenges for efficient inference. The memory wall, the growing processor-memory speed disparity, remains a critical bottleneck for LLM. Process-In-Memory (PIM) architectures overcome limitations by co-locating compute units with memory, leveraging 5-20$\times$ higher internal bandwidth and enabling greater energy efficiency than GPUs. However, existing PIMs struggle to balance flexibility, performance, and cost-efficiency for LLMs' dynamic memory-compute patterns and operator diversity. DRAM-PIM suffers from inter-bank communication overhead despite its vector parallelism. SRAM-PIM offers sub-10ns latency for matrix operation but is constrained by limited capacity. This work introduces CompAir, a novel PIM architecture that integrates DRAM-PIM and SRAM-PIM with hybrid bonding, enabling efficient linear computations while unlocking multi-granularity data pathways. We further develop CompAir-NoC, an advanced network-on-chip with an embedded arithmetic logic unit that performs non-linear operations during data movement, simultaneously reducing communication overhead and area cost. Finally, we develop a hierarchical Instruction Set Architecture that ensures both flexibility and programmability of the hybrid PIM. Experimental results demonstrate that CompAir achieves 1.83-7.98$\times$ prefill and 1.95-6.28$\times$ decode improvement over the current state-of-the-art fully PIM architecture. Compared to the hybrid A100 and HBM-PIM system, CompAir achieves 3.52$\times$ energy consumption reduction with comparable throughput. This work represents the first systematic exploration of hybrid DRAM-PIM and SRAM-PIM architectures with in-network computation capabilities, offering a high-efficiency solution for LLM.

\end{abstract}




\maketitle

\input{Sections/intro-v2}
\input{Sections/background}
\input{Sections/observation}
\input{Sections/architecture}
\input{Sections/noc}
\input{Sections/programming}
\input{Sections/evalutionv2}
\input{Sections/discussion}
\input{Sections/related}

\section{Conclusion}

The paper introduces CompAir, a novel architecture with hybrid bonding of DRAM-PIM and SRAM-PIM to address the challenges of efficient LLM inference. In CompAir, the CompAir-NoC enables low-cost non-linear operations and efficient collective communication, further optimizing the performance. Finally, we propose a hierarchical ISA to ensure programmability across hybrid PIM systems. CompAir represents the systematic exploration of scalable hybrid PIM architectures, offering an energy-effective and efficient solution for LLM inference. 



\bibliographystyle{ACM-Reference-Format}
\bibliography{refs, model, noc}

\end{document}

%% file: Sections/intro-v2.tex
\section{Introduction}

\begin{figure}[htbp]
  \centering
  \includegraphics[width=\linewidth, trim=0cm 0cm 0cm 0.2cm, clip]{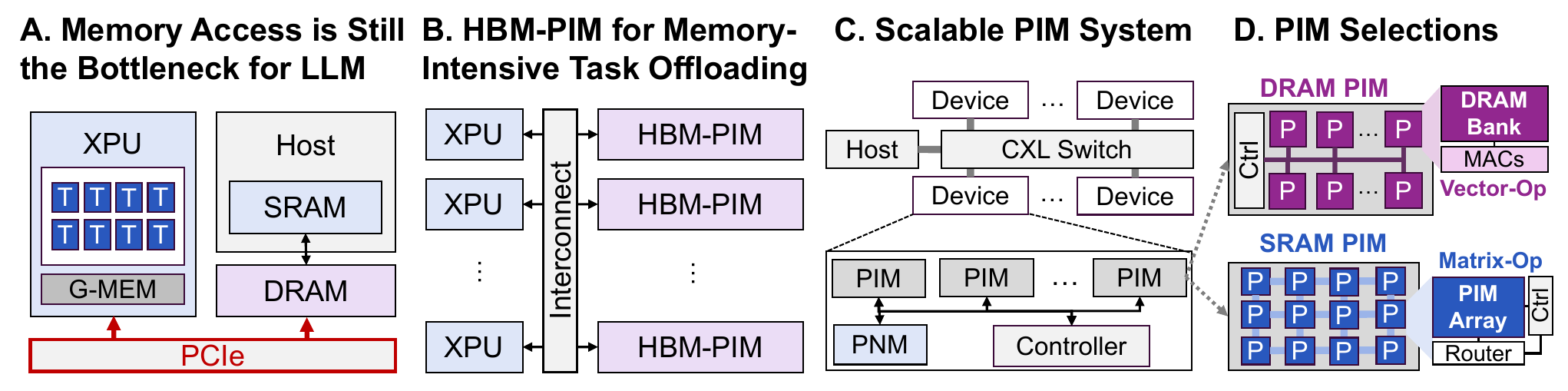}
  \centering
  \small
  \begin{tabular}{ccc}
    \toprule
    PIM Configuration & Energy Cost & Performance \\
    \midrule
    XPU+HBM-PIM~\cite{IANUS,AttAcc} & \textcolor[rgb]{0.5,0,0}{High} & \textcolor[rgb]{0,0.5,0}{High} \\
    Only DRAM-PIM~\cite{gu2025cent,lolpim} & \textcolor[rgb]{0,0.5,0}{Low} & \textcolor[rgb]{0.5,0.5,0}{Medium} \\
    Only SRAM-PIM~\cite{Tu2023TenserCIM,AIG-CIM} & \textcolor[rgb]{0,0.5,0}{Low} $\times$ \textcolor[rgb]{0.5,0,0}{Too Many Macros} & \textcolor[rgb]{0,0.5,0}{High} \\
    Hybrid PIMs \textbf{(Our Goal)} & \textcolor[rgb]{0,0.5,0}{Low} & \textcolor[rgb]{0,0.5,0}{High} \\
    \bottomrule
  \end{tabular}
  \caption{The motivation of CompAir.}
  \label{fig:motivation}
\end{figure}

The rapid advancement of Large Language Models (LLMs)\cite{llama2,deepseek,tranformer} is driving transformative changes, but their massive parameters and computational demands lead to prohibitive costs. For instance, LLaMA-65B inference consumes $2\times10^3$ Joules per response (batch size 512, 32 V100 GPUs\cite{LLMEnergy}) and commercial deployments like ChatGPT incur daily inference costs of approximately \$7 million\cite{LLMmoney}. Moreover, the scaling law\cite{scale} dictates a continual increase in model size, exacerbating computational bottlenecks. A fundamental constraint in LLM inference is the memory wall, where the growing disparity between processor speed and memory access\cite{MemoryWall} severely limits efficiency\cite{MemoryWallAI}. LLM inference architectures (Fig. \ref{fig:motivation}A) typically compose of XPUs (tensor accelerators like GPUs\cite{A100} and TPUs\cite{TPU, TPUv4}) interconnected with DRAM through PCIe (about 64 GB/s\cite{PCIe}) suffering from extreme data transfer bottleneck. For OPT-66B\cite{OPT66B}, PCIe transfers contribute 90\% of inference latency\cite{AffordableNDPDIMMs}. While model compression methods like  quantization\cite{awq}, pruning\cite{powerinfer, DejaVu}, and low-rank adaptation\cite{lora} alleviate bandwidth constraints, they fail to break the fundamental memory bottleneck.

Process-In-Memory (PIM) architectures offer a promising solution to overcome memory bottlenecks by leveraging the high internal memory bandwidth: 6.7$\times$ higher than the external bandwidth in UPMEM\cite{UPMEM} and 16$\times$ in AiM\cite{AiM}. PIM enables in situ data processing that reduces energy consumption and improves throughput. Several memory technologies have embraced this architecture, including DRAM\cite{Rowclone, MIMDRAM}, Non-Volatile Memory (NVM)\cite{PRIME}, NAND Flash\cite{sflash}, and SRAM\cite{CompCache, VSPIM}. Among these, DRAM-PIM\cite{UPMEM, PathfindingUPMEM, FIMDRAM} and SRAM-PIM\cite{CIM-14-2-isscc25, Tu2023TenserCIM, SRAMpimISSCC23} stand out as promising candidates for real-world deployment due to their high endurance (larger than $10^{16}$), process compatibility, and scalability\cite{PIMSurvey}. Recent advances have shown the potential of offloading memory-bound operations to high-bandwidth memory (HBM), such as Generalized Matrix-Vector Multiplication (GeMV), yielding inference performance gain by alleviating bandwidth limitations\cite{NeuPIMs, IANUS, AttAcc} (Fig. \ref{fig:motivation}B). Yet, LLM inference remains energy inefficient. Both XPUs and HBMs are notorious for high power consumption, prompting a search for alternative architectures. Two key directions have emerged: \textit{(i)} design of fully XPU-free PIM systems, and \textit{(ii)} the use of more energy-efficient memory technologies than HBM.

\begin{figure*}[htbp]
  \centering
  \includegraphics[width=0.95\linewidth]{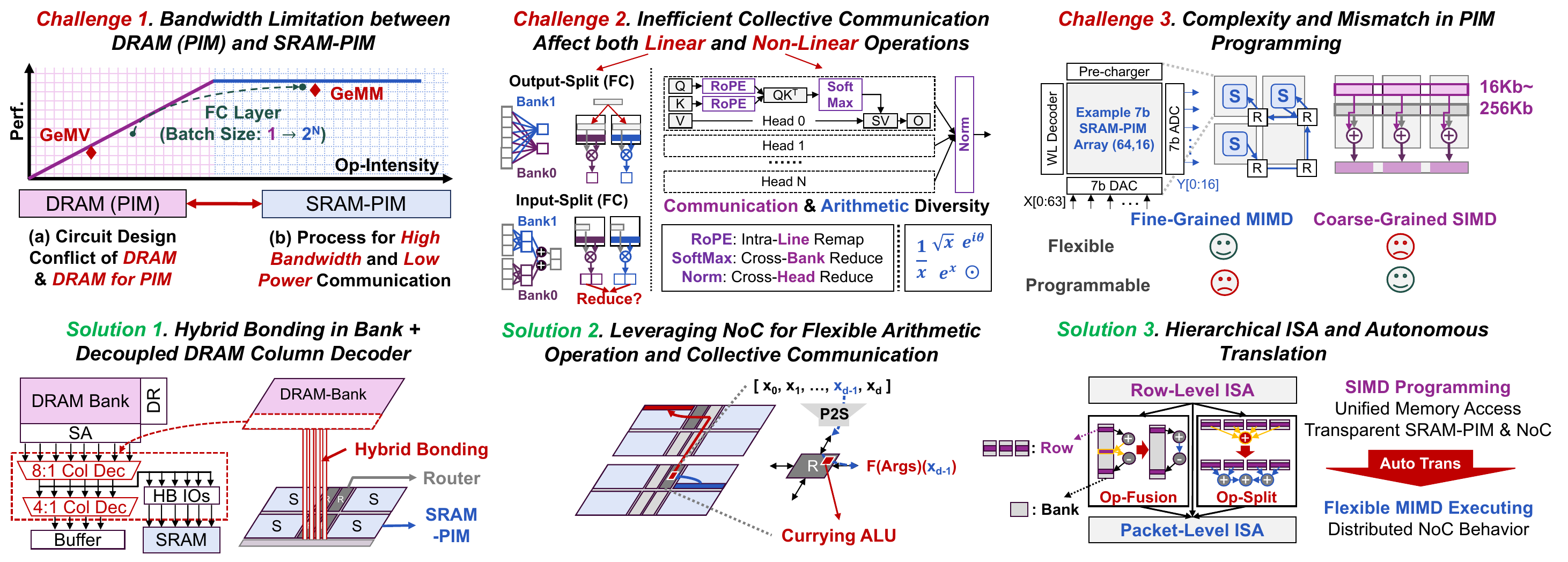}
  \caption{Key challenges in hybrid PIM for LLM and corresponding solutions in CompAir.}
  \label{fig:sol}
\end{figure*}

In response, recent work has explored fully PIM-based architectures (Fig. \ref{fig:motivation}C)\cite{gu2025cent,lolpim}. XPU-free PIM systems incorporate Compute Express Link (CXL) to minimize inter-device communication latency and scale efficiently. To support complex non-linear operations essential for LLMs, centralized CPUs and massive dedicated non-linear units are incorporated as a Process-Near-Memory (PNM) module in the CXL controller, enabling area-efficient implementation of operations such as RoPE and Softmax\cite{gu2025cent}. 


\textcolor{black}{The dynamic memory-compute behavior of LLMs, driven by variable batch sizes and sequence lengths, remains memory substrate selection a critical challenge} (Fig. \ref{fig:motivation}D). SRAM-PIM excels at low-latency (<10 ns) matrix operations with high efficiency in workloads with substantial weight reuse. However, it suffers performance degradation for write-intensive operations (e.g., attention), and its small macro size\cite{PIM4AI} imposes excessive power and area overhead when scaled to LLM-scale workloads (Fig. \ref{fig:sram_vs_dram}). Conversely, GDDR6-based DRAM-PIM achieves superior cost and energy efficiency over XPU+HBM-PIM\cite{gu2025cent,lolpim}, but faces compute-bound bottlenecks at large batch sizes\cite{UPMEMbanchmark} and high inter-bank communication overhead during long-context inference\cite{UMPIM} (Fig. \ref{fig:acc}). To address these constraints, we present \textbf{\textit{hybridizing DRAM-PIM and SRAM-PIM for memory-bounded and compute-bounded tasks respectively}}, to achieve efficient LLM inference.

However, hybridizing DRAM-PIM and SRAM-PIM into a unified system introduces several fundamental challenges. As illustrated in Fig. \ref{fig:sol}, we identify three critical issues and propose targeted solutions to address them that underpin the CompAir architecture:




\textbf{Challenge 1: Bandwidth Bottleneck.} The data movement between DRAM and SRAM is constrained by interconnect bandwidth at two levels: \textit{(i)} To accommodate more logic, modern DRAM-PIMs place compute units outside column decoders\cite{AiM,Newton,FIMDRAM}. While improving logic density, it significantly reduces accessible bit width. For instance, 96.8\% DRAM \textcolor{black}{read-out bandwidth} is sacrificed in Newton\cite{Newton}, which is insufficient to feed SRAM-PIM's high-throughput demands. \textit{(ii)} Process constraints require separate dies for DRAM and SRAM, where the limited inter-die bandwidth\cite{BeachfrontLimitations} exacerbates the bottleneck.


\textbf{\textit{Solution 1: Hybrid Bonding with Decoupled Column Decoder.}} For Challenge 1(i), we propose a decoupled column decoder in DRAM-PIM that simultaneously maintains standard DRAM functionality while enabling high-bandwidth data access tailored for SRAM-PIM. For Challenge 1(ii), we adopt hybrid bonding (HB)\cite{HybridBondingReview, HBA} with area-matched SRAM-PIMs and DRAM-PIM bank. This cross-die alignment supports distributed, high-throughput communication, substantially alleviating the interconnect bottleneck.


\textbf{Challenge 2: Inefficient Fine-Grained Collective Communication}. \textcolor{black}{Inter-bank communication remains a critical bottleneck for LLM inference in contemporary DRAM-PIMs}\cite{UMPIM}. The bottleneck manifests in two critical ways: \textit{(i) Non-linear overhead.} Non-linear operations require fine-grained data rearrangement\cite{tranformer}. Our profiling reveals that in long-context scenarios, communication for non-linear computation can account for up to 25\% of total latency. \textit{(ii) Suboptimal reduce-avoidance strategies.} To avoid inter-bank collective communication overhead, prior DRAM-PIM solutions restrict output-split mapping (shown in the left of Fig. \ref{fig:sol} Challenge 2) for fully-connected (FC) layers \cite{gu2025cent, lolpim}. However, our experiments shows that this strategy often leads to suboptimal execution.


\textbf{\textit{Solution 2: Low-Cost In-transit Computable NoC.}} To address both issues, we introduce CompAir-NoC, a novel computable Network-on-Chip (NoC). At its core is the Currying ALU, a low-cost arithmetic logical unit embedded within routers. This design enables non-linear operations to be performed in-transit, avoiding costly data movement to centralized PNM, while simultaneously enabling efficient inter-bank collective communications. By integrating computation into the communication fabric, CompAir-NoC offers a low-cost, distributed mechanism that both accelerates non-linear functions and inter-bank collective communication, significantly improving system-wide performance.


\textbf{Challenge 3: Programming Mismatch.} Hybrid PIM architectures combines DRAM-PIM and SRAM-PIM, which inherently adopt distinct execution models. DRAM-PIM employs SIMD executions with centralized control and shared instruction contexts across all banks\cite{AiM, UPMEM, UPMEMbanchmark}, while SRAM-PIM utilizes an MIMD paradigm with distributed controllers and private instruction contexts per bank for flexibility\cite{Tu2023TenserCIM, AIG-CIM}. \textcolor{black}{However, extending MIMD across all banks imposes substantial programming complexity and incurs significant area cost overhead due to private instruction buffer}, up to 20\% of the logic die\cite{HBA}. This architectural disparity poses a fundamental challenge to the programmable hybrid PIM systems.

\textbf{\textit{Solution 3: Hierarchical ISA with Automated Translation.}} To reconcile programmability with architectural heterogeneity, we propose a two-level ISA abstraction with autonomous translation, combining the simplicity of SIMD programming with the flexibility of MIMD execution. At the \textbf{row-level}, we retain a unified SIMD instruction interface and memory access patterns for ease of programming. At the \textbf{packet-level}, we introduce programmable routing behaviors that enable MIMD-like execution. A key innovation lies in overcoming the inefficiency of directly mapping SIMD-style programs to MIMD networks, which risks underutilizing the NoC’s potential for fine-grained parallelism. Inspired by AI compilers\cite{TVMopfusion}, we design a novel instruction-level operator fusion and split mechanism, enabling automated NoC path synthesis by analyzing address dependencies across row-level instructions.

A detailed technical analysis of these observations is provided in Section \ref{sec:observation}. The key contributions of this work are listed below:



\begin{enumerate}
    \item We introduce CompAir, a hybrid PIM architecture integrating DRAM-PIM, SRAM-PIM with hybrid bonding to achieve energy-efficient and scalable LLM inference. (Section \ref{sec:arch})
    \item We develop CompAir-NoC, a low-latency NoC with a novel ALU to achieve efficient and low-cost non-linear operations and in-network collective communication. (Section \ref{sec:noc})
    \item We design a novel hierarchical ISA, overcoming the programming issues and enabling transparent and scalable execution across the hybrid PIM systems. (Section \ref{sec:prog})
\end{enumerate}


To the best of our knowledge, \textit{CompAir is the first architecture that systematically addresses PIM hybridization with fundamentally different programming models}, achieving a balanced trade-off among performance, energy efficiency and programmability for LLM. CompAir achieves 1.83-7.98$\times$ prefill and 1.95-6.28$\times$ decode improvement over the state-of-the-art fully PIM architecture. Compared to the hybrid A100 and HBM-PIM system, CompAir achieves 3.52$\times$ energy consumption reduction with comparable throughput.

%% file: Sections/observation.tex
\section{Background and Motivations}
\label{sec:observation}


\subsection{Transformer and LLM Model}
\label{sec:bgd_llm}

\begin{figure}[h]
  \centering
  \includegraphics[width=0.9\linewidth, trim=0cm 0.2cm 0cm 0cm, clip]{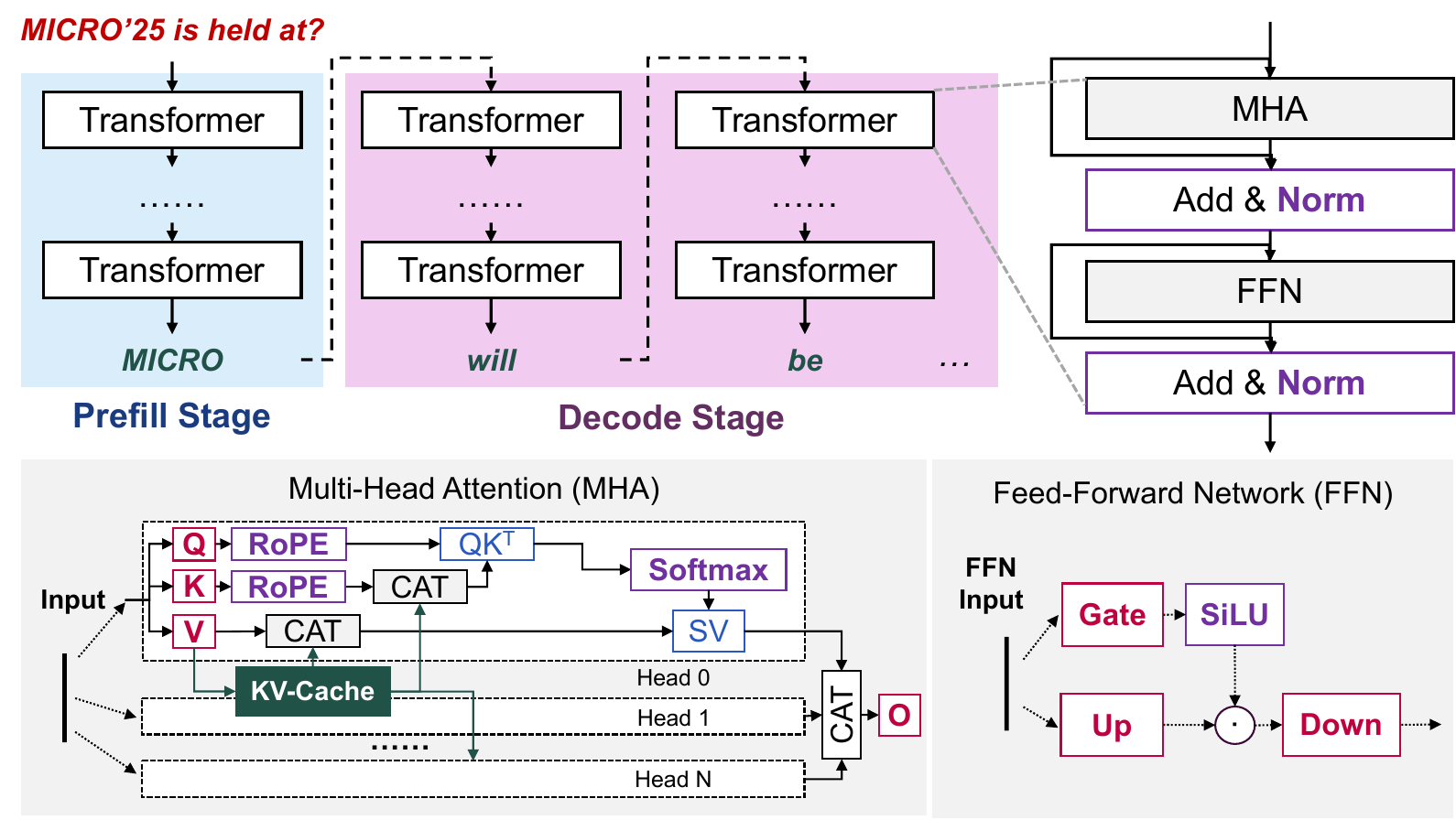}
  \caption{LLM framework (Llama2 model architecture). In MHA and FFN, we use the \textcolor{purple}{purple} blocks to represents the FC layers and \textcolor{violet}{violet} blocks to represent non-linear operations.}
  \label{fig:llm}
\end{figure}

Modern transformer-based LLMs (Fig. \ref{fig:llm}) operate in two distinct stages: the prefill stage, which processes all tokens using matrix-based operations, and the decoding stage, which sequentially generates tokens using vector-based operations. The Multi-Head Attention (MHA) initially projects inputs to queries (Q), keys (K), and values (V) via fully connected (FC) layers, followed by Rotary Positional Embedding (RoPE) \cite{tranformer} to encode relative positional information. Attention scores S are computed via the product of Q and K, normalized through a Softmax function, and used to weight V. A final linear projection O yields the attention output. To efficiently support long-context processing, LLMs adopt KV cache\cite{KVCacheSuvey}. For example, Llama2 maintains individual KV caches for each attention head. Downstream of attention, RMSNorm \cite{RMSNorm} normalizes activations, followed by a Feed-Forward Network (FFN) with parallel Up and Gate projections, gated via the SiLU non-linearity, and a Down projection that generates the final output.

\subsection{DRAM-PIM and SRAM-PIM Own Different Advantages in LLM Inference}
\label{ob_linear}

\begin{figure}[h]
  \centering
  \includegraphics[width=0.9\linewidth, trim=0cm 0cm 0cm 0.2cm, clip]{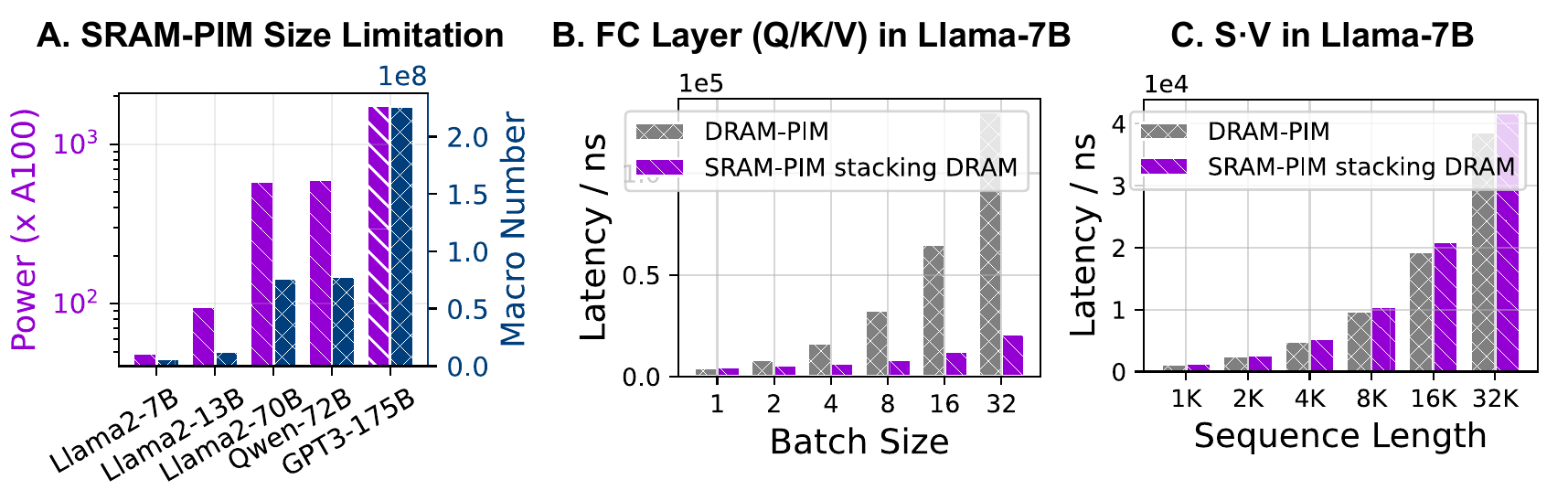}
  \caption{Comparison between DRAM-PIM\cite{AiM}, \textcolor{black}{pure} SRAM-PIM\cite{SRAMpimISSCC23} and \textcolor{black}{SRAM-PIM stacking DRAM} \textcolor{black}{in decoding}. (A) \textcolor{black}{Pure} SRAM-PIMs compute all FC layers in different models without weight reloading, the power and macro number are both unacceptable. (B) and (C) set four 8KB SRAM-PIM macros for each DRAM bank in Q/K/V and SV.}
  \label{fig:sram_vs_dram}
\end{figure}

While recent efforts have optimized linear operations within LLMs, hardware constraints remain a major determinant of inference efficiency. Specifically, DRAM-PIM and SRAM-PIM architectures exhibit complementary strengths and limitations. Using Llama2-7B as a case study, we compare their performance across batch sizes and sequence lengths as shown in Fig. \ref{fig:sram_vs_dram}.


\textcolor{black}{\textit{Pure SRAM-PIM designs are fundamentally impractical for LLMs.}} As demonstrated in Fig. \ref{fig:sram_vs_dram}A, implementing GPT3-175B solely with SRAM-PIM would require an infeasible number of macros and exceed the power consumption of an NVIDIA A100 GPU (assumed 300W) by three orders of magnitude even for only FC layers. \textcolor{black}{This indicates the importance of extending the DRAM bank for SRAM-PIM, which is the focus of the subsequent analysis and pure SRAM-PIM will not be taken into consideration, but then DRAM bandwidth becomes the critical bottleneck.} \textcolor{black}{One contemporary solution is to solve this problem by stacking DRAM on the logic die\cite{HBA}, so we further compare the performance of SRAM-PIM stacking DRAM and pure DRAM-PIM. In Fig. \ref{fig:sram_vs_dram}B, SRAM-PIM stacking DRAM} offers no advantage over DRAM-PIM due to overheads associated with frequent weight writes when batch=1. However, at batch size=32, \textcolor{black}{SRAM-PIM stacking DRAM} achieves a 6.3$\times$ speedup over DRAM-PIM, capitalizing on its superior weight reuse. This aligns with the expected shift from memory-bound (GeMV) to compute-bound (GeMM) behavior in Q/K/V projection as batch size grows. 


Unfortunately, this feature can not apply to all linear operators in LLM. In $QK^T$ and $SV$, the matrices ($K^T$ and $V$) are input-dependent and dynamically shaped by sequence length, making them unsuitable for SRAM-PIM due to poor matrix reusing opportunities. As Fig. \ref{fig:sram_vs_dram}C shows \textcolor{black}{SRAM-PIM stacking DRAM} underperforms DRAM-PIM for $SV$, reverting the system to a batch=1 regime in Fig. \ref{fig:sram_vs_dram}B.

In all, these findings demonstrate that SRAM-PIM can deliver performance gains far beyond DRAM-PIM for batched FC layers. But this benefit is non-uniform across LLM workloads, and is constrained by several factors including bandwidth, heat, and mapping schemes, which will be discussed in detail in section \ref{sec:arch}.



\subsection{Non-Linear Operations Cannot be Ignored}
\label{ob_non_linear}

\begin{figure}[h]
  \centering
  \includegraphics[width=0.9\linewidth, trim=0cm 0.6cm 0cm 0cm, clip]{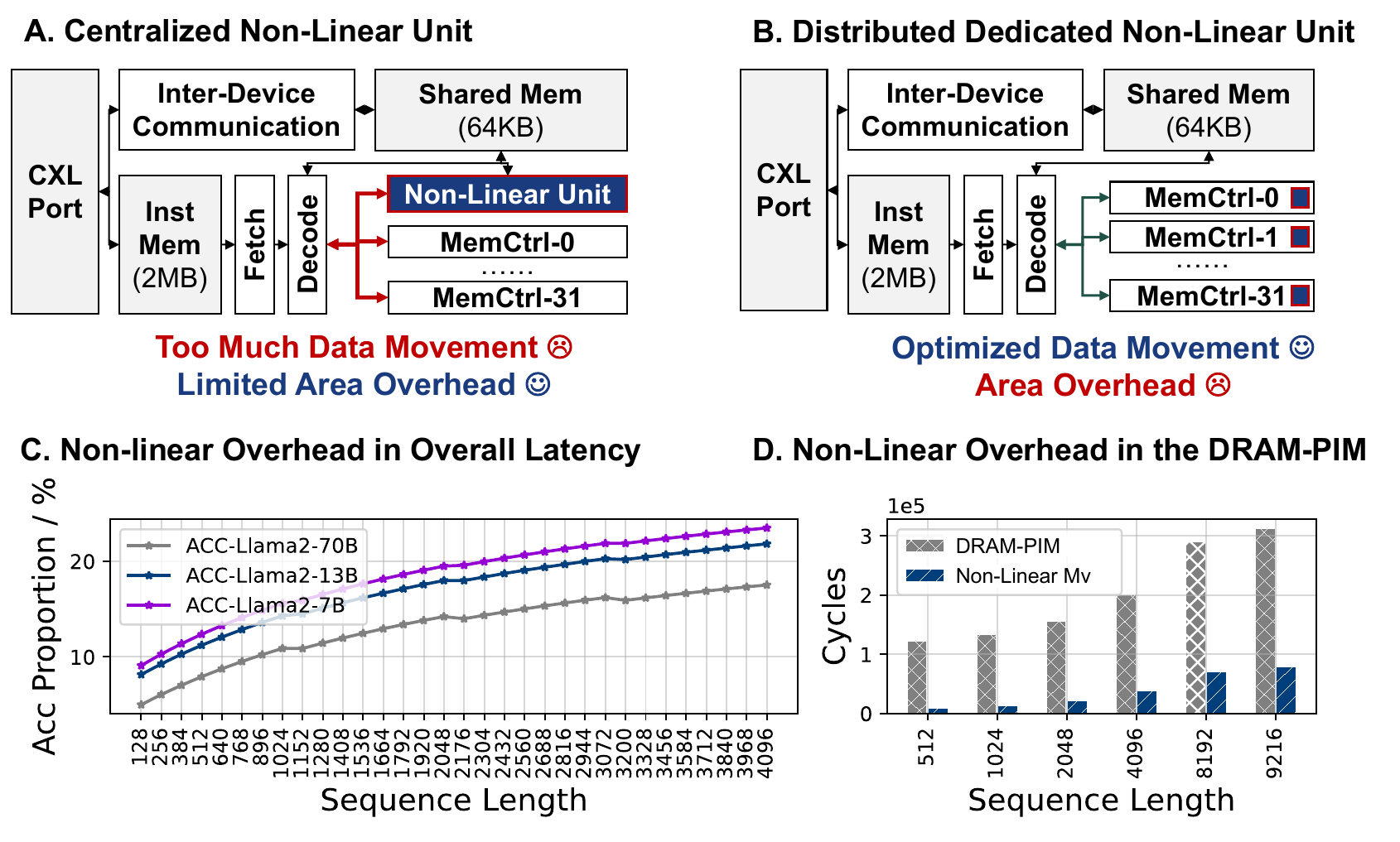}
  \caption{Non-linear overhead is not negligible. \textcolor{black}{(A) Having all channels share the same NLU results in a lot of data movement between the NLU and each channel. (B) Tailoring NLU within each channel or bank incurs an area cost.} (C) The proportion of non-linear operation in the transformer. (D) Extra data movement for non-linear operations in DRAM-PIM\cite{gu2025cent}.}
  \label{fig:acc}
\end{figure}

While prior research has predominantly focused on optimizing linear operations, non-linear operations are becoming a significant bottleneck in long-context LLM inference. Three strategies are commonly employed to address non-linear computation: \textit{(i)} Offloading non-linear operations to GPUs\cite{AttAcc} or NPUs with dedicated non-linear units(NLU)\cite{NeuPIMs}. \textit{(ii)} Centralized non-linear units and CPUs located outside of the DRAM-PIM channels\cite{gu2025cent, lolpim} \textcolor{black}{(Fig. \ref{fig:acc}A)}. \textit{(iii)} Distributed non-linear units near each bank \textcolor{black}{(Fig. \ref{fig:acc}B)}.

Method (i) depends on high-performance GPUs, while CENT [1] shows method (iii) faces challenges from diverse non-linear operators in LLMs. Non-linear units require significant area: 4.4mm² (7nm)\cite{gu2025cent} - 4× larger than a 32MB DRAM bank - and up to 30\% of FPGA resources\cite{AiM}. Thus, method (ii) has been typically preferred under area/power constraints.




Yet, the increasing adoption of long-context reasoning in LLMs, supporting up to 128K tokens\cite{reasoning, qwen2}, is reshaping the computational paradigm of modern AI systems. Our analysis \textcolor{black}{based on pure DRAM-PIM\cite{gu2025cent} with centralized NLU} demonstrates a significant shift in performance bottlenecks. At a 4K token sequence length, non-linear operations (such as Softmax, whose latency scales with the sequence length) account for about 20\% of the total execution time of the transformer block (Fig. \ref{fig:acc}C). Critically, these non-linear operations impose substantial communication costs due to the required reduction and broadcasting across memory banks and channels. \textcolor{black}{Fig. \ref{fig:acc}D shows that in long-context scenarios, DRAM-PIM non-linear computation overheads can exceed 25\% of total inference time. This contradicts the assumption that non-linear ops are secondary, revealing them as key bottlenecks at scale. New architectural non-linear support is needed for efficient LLM inference.}

%% file: Sections/architecture.tex
\section{CompAir Architecture}
\label{sec:arch}

\begin{figure}[h]
  \centering
  \includegraphics[width=0.95\linewidth, trim=0cm 0cm 0cm 0.35cm, clip]{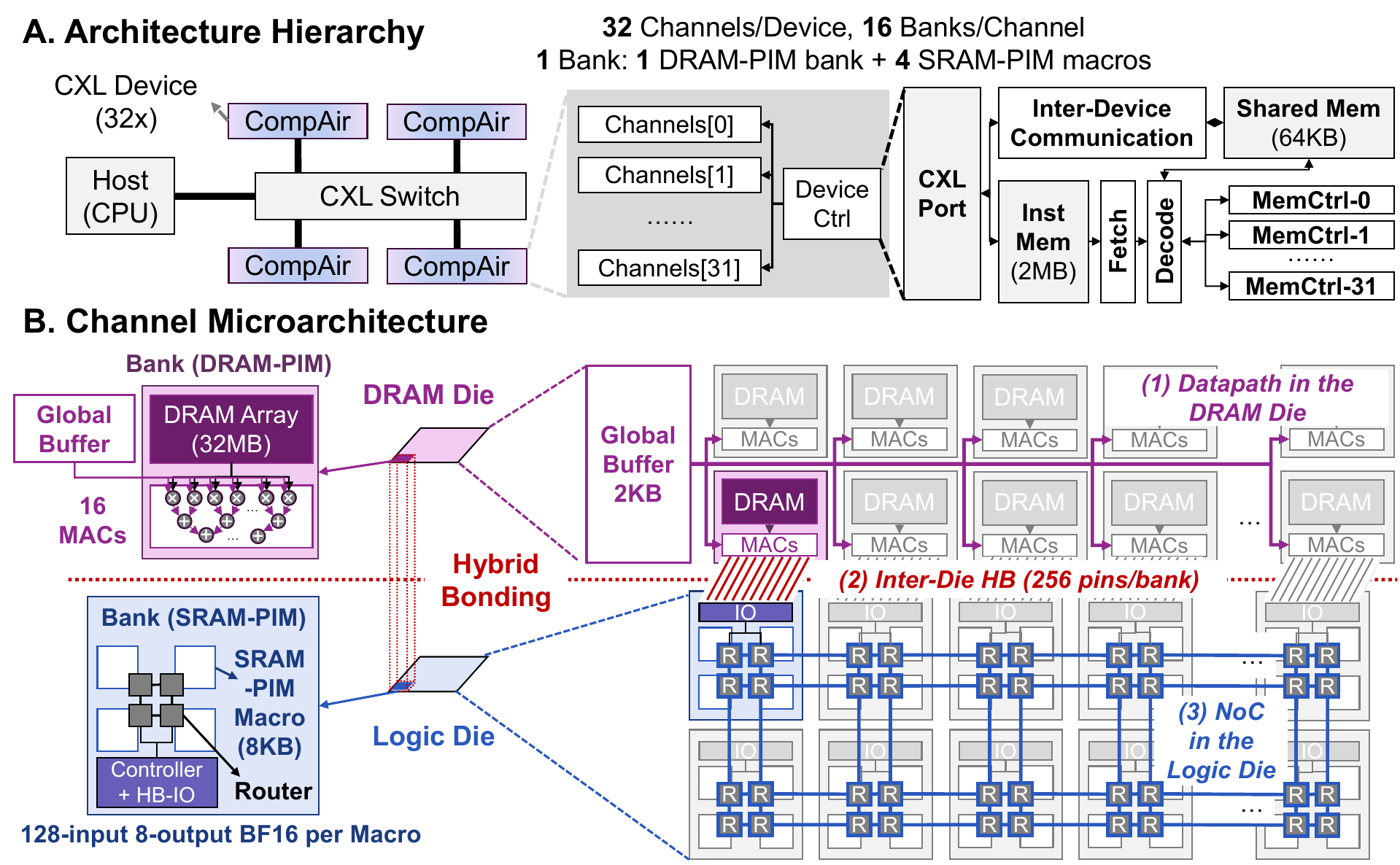}
  \caption{Architecture of CompAir.}
  \label{fig:harch}
\end{figure}

In section \ref{sec:observation}, we identified key performance bottlenecks in existing LLM-oriented DRAM-PIM \textcolor{black}{and SRAM-PIM stacking DRAM} architectures, motivating our proposal of a hybrid PIM system that integrates both DRAM-PIM and SRAM-PIM technologies. Fig. \ref{fig:harch} presents the architecture of CompAir.


This section focuses on the challenges and innovations underpinning the hybrid DRAM-PIM and SRAM-PIM integration. In CompAir, we adopt CLX.io and CXL.mem in the CXL protocols to enable scalable communication, achieving up to 29.44 GB/s collective cross-device broadcast/reduce operations and 53.5 GB/s for point-to-point transfers. A total of 32 PIM-enabled devices are connected via the CXL switch (Fig. \ref{fig:harch}A)\cite{CXL}. Each device hosts a lightweight controller that contains instruction and shared memory. Unlike prior designs\cite{lolpim, gu2025cent}, CompAir's device controllers are only responsible for instruction issuance and do not contain the non-linear execution units. Within each device, the controller controls 32 independent memory channels, each containing 16 CompAir banks composed of tightly integrated DRAM-PIM and SRAM-PIM with hybird bonding (Fig. \ref{fig:harch}B). \textcolor{black}{The design integrates a DRAM die with DRAM-PIM and a logic die with SRAM-PIM macros, hybrid bonding I/Os, and a NoC. Each DRAM-PIM bank includes a 16-input BF16 MAC unit, with inter-bank communication through a global buffer. In the logic die, each SRAM-PIM bank comprises four SRAM-PIM macros and four routers. Routers in the logic die are interconnected as the NoC. DRAM-PIM and SRAM-PIM banks are paired 1:1 across dies, communicating through 256 bonds per bank.}


To substantiate our design, we address three key issues for DRAM-PIM and SRAM-PIM integration, guided by existing platforms (AiM\cite{AiM, gu2025cent} and a fabricated SRAM-PIM\cite{SRAMpimISSCC23} without modifications at the circuit level). These challenges include integration granularity (section \ref{sec:hybird_grain}), hardware specification and feasibility (section \ref{sec:arch_hw_issue}), and mapping constraints (section \ref{sec:arch_sw_issue}). Finally, we demonstrate that targeted micro-architectural refinements to DRAM-PIM can yield substantial end-to-end performance gains (section \ref{sec:modify}).


\subsection{Why Intra-Channel Hybridization?}
\label{sec:hybird_grain}

A central design question is how can we achieve efficient heterogeneous integration of DRAM-PIM and SRAM-PIM to fully exploit their complementary advantages. We explore three possible integration schemes: \textit{(i)} inter-device hybridization, \textit{(ii)} inter-channel hybridization, and \textit{(iii)} intra-channel hybridization.


\begin{figure}[h]
  \centering
  \includegraphics[width=0.92\linewidth, trim=0cm 0.6cm 0cm 0.4cm, clip]{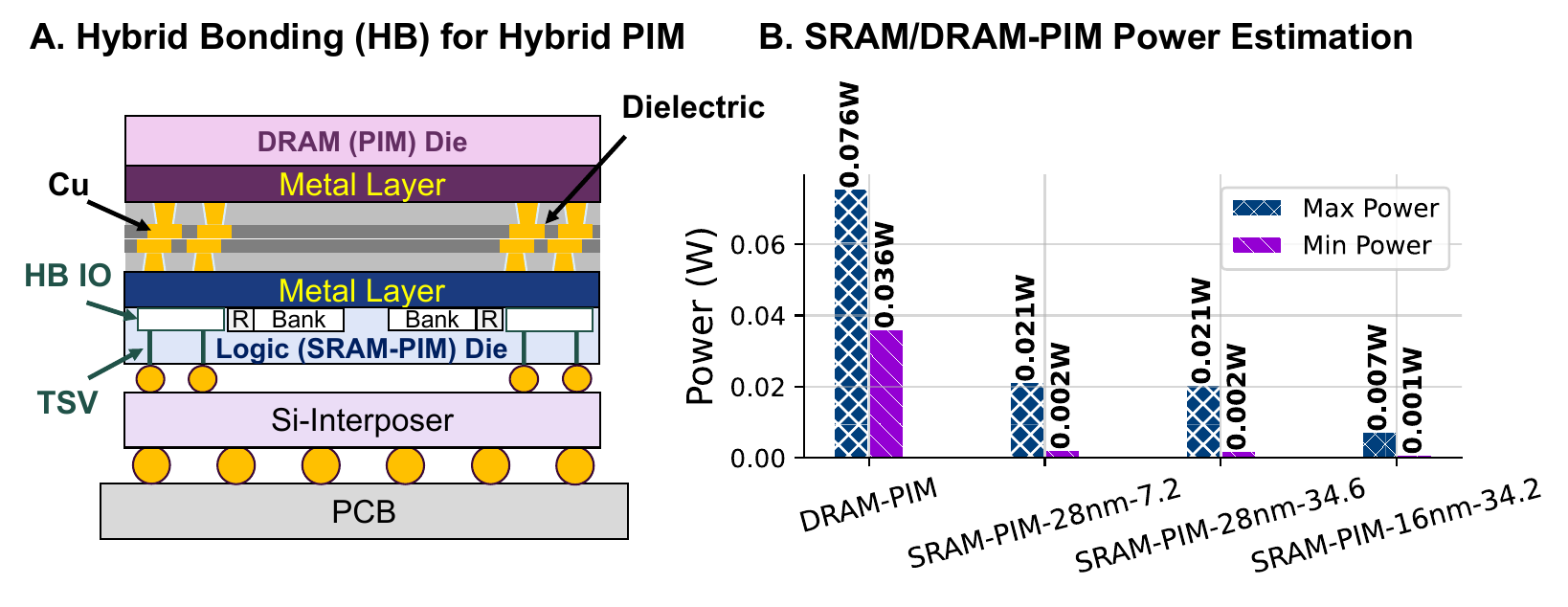}
  \caption{Hardware issues of CompAir. (A) The illustration of hybrid bonding. (B) The estimated power of one DRAM-PIM bank and SRAM-PIMs with 8KB\cite{ISSCC-34-6, SRAMpimISSCC23, ISSCC-34-2}.}
  \label{fig:hw}
\end{figure}
\vspace{-0.25cm}

The first two schemes have strong limitations due to limited bandwidth. While weight preloading can reduce the data movement to SRAM-PIM, it is impractical at scale. Even computing only Q/K/V projections in a 70B model\cite{llama2} would require 192GB of SRAM, far exceeding current SRAM macro sizes (224kb in BF16\cite{ISSCC25-14-4}. Therefore, weight reloading is inevitable for SRAM-PIM, and we choose intra-channel hybridization to guarantee the bandwidth between SRAM and DRAM as shown in Fig. \ref{fig:harch}B. Taking AiM\cite{AiM} as an example, the internal bandwidth of a single channel of DRAM is 512GB/s, while external I/O bandwidth is limited to 32GB/s. Even a compact 128-input, 8-output INT8-precision SRAM-PIM, operating at 16ns latency, demands 64GB/s to remain fully utilized. Moreover, the beachfront problem\cite{BeachfrontLimitations} indicates that there is not enough edge on the die for all fibers to connect by going vertically. To resolve this, CompAir leverages hybrid bonding (HB)\cite{HybridBondingReview} (Fig. \ref{fig:hw}A) for 3D integration, stacking SRAM-PIM macros under each DRAM-PIM bank. HB achieves bonding densities of 10K-100K interconnects per $mm^2$ density with an energy cost of just 0.05-0.88pJ/b, which is over 200$\times$ more efficient than off-chip HBM\cite{ISSCC22HB}. However, this architecture demands careful analysis at both hardware and software levels. Two fundamental questions remain: \textit{(i)} Is this heterogeneous hybridization feasible under current hardware constraints\textcolor{black}{(Sections \ref{sec:arch_hw_issue})}?  \textit{(ii)} What are the mapping implications for efficient DRAM-PIM and SRAM-PIM collaboration\textcolor{black}{(Sections \ref{sec:arch_sw_issue})}?

\begin{figure*}[h]

    \begin{minipage}{\linewidth}
        \centering
	\includegraphics[width=0.9\linewidth, trim=0cm 0.45cm 0cm 0.1cm, clip]{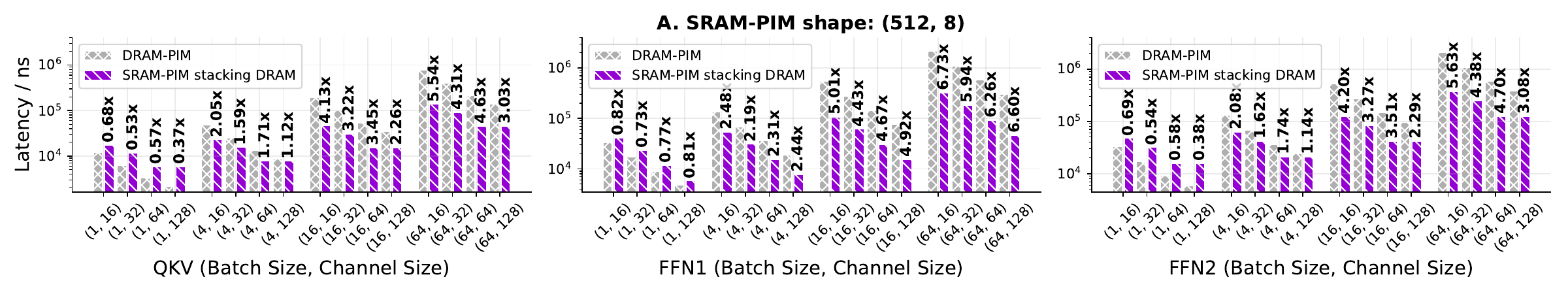}
    \end{minipage}
    \qquad
	
    \begin{minipage}{\linewidth}
	   \centering
	   \includegraphics[width=0.9\linewidth, trim=0cm 0.45cm 0cm 0.1cm, clip]{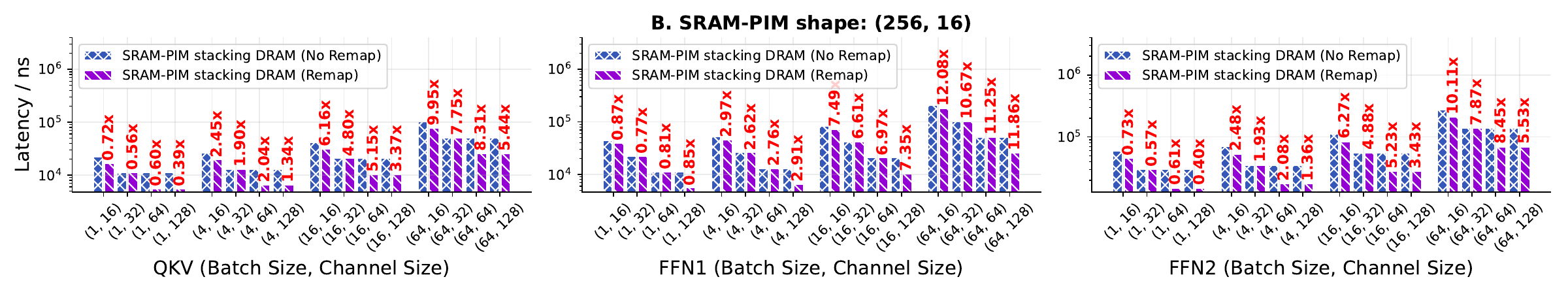}
    \end{minipage}

  \caption{\textcolor{black}{Pure DRAM-PIM} and \textcolor{black}{SRAM-PIM stacking DRAM} performance with Llama2-13B. Labeled numbers are improvement relative to \textcolor{black}{pure DRAM-PIM}.}
  \label{fig:sram_vs_dram_deep}
\end{figure*}

\subsection{Area and Power Issue}
\label{sec:arch_hw_issue}


Piror work\cite{HBA} has illustrated that centralized IO controllers can lead to severe performance loss, and therefore, we need to establish a local pairing between DRAM-PIM banks and SRAM-PIMs. This requires a matching in area between the two levels. It also needs to be ensured that the extra power consumption introduced by the SRAM-PIM is acceptable, otherwise this will meet the heating issue.

For the area issue, the 1ynm 32MB bank of an existing DRAM-PIM has an area of around 1$mm^2$\cite{AiM}, while a 28nm 8KB SRAM-PIM macro occupies 0.136$mm^2$\cite{ISSCC25-14-6}. Therefore, integrating four 8KB SRAM-PIM macros under each DRAM-PIM is a feasible specification. This issue will also be analyzed in detail in section \ref{sec:evaluation}.

SRAM-PIM architectures have demonstrated order-of-magnitude gains in energy efficiency over conventional neural processing units, achieving >30 TFLOPS/W\cite{ISSCC25-14-5, ISSCC-34-6, SRAMpimISSCC23, ISSCC-34-2}, compared to <5 TFLOPS/W for most NPUs\cite{AIAccSurvey}. This compelling efficiency advantage motivates our selection of SRAM-PIM as the foundational matrix computation unit in CompAir. In Fig. \ref{fig:hw}B, we analyze the power consumption of DRAM-PIMs running GPT3-175B workloads\cite{gpt3}, observing a power consumption of 0.036W to 0.076W per bank. In contrast, 8KB SRAM-PIMs consume merely 0.022W\cite{ISSCC-34-6, SRAMpimISSCC23, ISSCC-34-2}, which can drop further to 0.002W in low-voltage mode. Given that DRAM-PIM and SRAM-PIM operations are temporally decoupled, the incremental power overhead of incorporating SRAM-PIM is negligible while delivering substantial performance benefits.


\subsection{Organization and Mapping Issue}
\label{sec:arch_sw_issue}

\textcolor{black}{Section \ref{sec:arch_hw_issue} estimates the suitable size of SRAM-PIMs: one DRAM-PIM bank corresponds to four 8KB SRAM-PIMs shown in Fig. \ref{fig:harch}. Each SRAM-PIM macro is a 128-input 8-output BF16 matrix multiplication unit. In CompAir, SRAM-PIM is responsible for calculating FC layers with scenarios dedicated modification.} The 512GB/s of internal bandwidth mentioned in the previous section is averaged over each bank at 32GB/s with 256-bit width. The data rate of hybrid bonding can achieve 6.4Gbps\cite{TSVbw}, which can fully meet the DRAM bandwidth requirements. 

Our analysis of mapping strategies for DRAM-PIM and SRAM-PIM architectures reveals key distinctions. In scalable DRAM-PIM systems, matrix multiplications are typically distributed across banks to exploit memory parallelism. Input-split introduces inter-bank reduction overheads, which is limited by the bandwidth of the global buffer and requires serializing the access of the DRAM banks\cite{AiM, gu2025cent, lolpim, Newton}. Consequently, output-split becomes the predominant DRAM-PIM mapping approach, though it demands extensive input vector broadcasting and creates an \textbf{extreme dimensional imbalance in the FC layers (long input vectors versus short outputs per bank)}. \textcolor{black}{When SRAM-PIM performs matrix multiplication, DRAM is responsible for fetching input to the SRAM-PIM and writing results back.} Therefore, the shape imbalance intensifies the DRAM-to-SRAM data movement pressure. SRAM-PIM favors balanced input-output mappings, where bandwidth demand is minimized when dimensions of inputs and outputs are similar for a given MAC count, according to the mean value inequalities.


To quantify these effects, we examine two configurations of four SRAM-PIM macros: (512,8)\footnote{The SRAM-PIMs are all configured with 128-inputs-8-outputs, so (512,8) refers to extending 4 SRAM-PIMs in the input dimension into a 512-input-8-output matrix unit.} and (256,16) \textcolor{black}{in different batch sizes and CompAir channels}. In Llama2-13B, each bank processes Q/K/V weight sized 5120$\times$10 \textcolor{black}{under the output-split mapping when 16$\times$32 banks are used in total (channel=32). SRAM-PIM retains weights across batches as much as possible, requiring DRAM-SRAM transfers only for input/output per inference and weight per reloading.}


In Fig. \ref{fig:sram_vs_dram_deep}A, \textcolor{black}{SRAM-PIM stacking DRAM deliver significant performance gains in both Q/K/V projection and FFN, with gains increasing alongside batch size, highlighting their superior data reuse efficiency. Fig. \ref{fig:sram_vs_dram_deep}B evaluates the trade-off in the (256,16) configuration. Although splitting along the input introduces modest reduction overheads, this layout substantially reduces DRAM-to-SRAM bandwidth stress, often yielding better overall performance than (512,8). When input-split mapping are also adopted (2560$\times$20 for each bank, channel=32)}, this reorganization consistently outperforms the pure output-split approach. These findings lead to two critical insights: (i) SRAM-PIM can lead to better performance in compute-bounded GeMM than DRAM-PIM. (ii) SRAM-PIM and DRAM-PIM have distinct mapping requirements for optimal performance. In hybrid PIM systems, efficient inter-bank reduction becomes critical.



However, the better mapping relies on efficient inter-bank reduction. Our detailed solution will be presented in section \ref{sec:noc}.

\subsection{DRAM-PIM Reorganizing}
\label{sec:modify}

\textcolor{black}{The design and synthesis of DRAM-PIMs depend on access to industrial PDKs, so we adopt AiM\cite{AiM} and its derivative designs\cite{gu2025cent, Newton} in the previous sections. However, CompAir architecture introduces new opportunities to rethink DRAM-PIM organization.}


Section \ref{sec:arch_sw_issue} identifies DRAM \textcolor{black}{read-out bandwidth} as the primary bottleneck in DRAM-SRAM interactions. This stems from current DRAM-PIM designs placing compute logic outside the column decoder to maximize logic integration\cite{PIM4AI}. Newton\cite{Newton} employs a 32:1 multiplexer for column selection, striking a balance between DRAM access and compute efficiency. This multiplexer is dubbed as column decoder. For a 1KB-wide DRAM array, single-row full-bitline access incurs excessive bandwidth overhead and restricts fine-grained memory operations. \textcolor{black}{Therefore, only 32B are typically accessed per operation, sufficient for traditional DRAM-PIM, but restrictive for hybrid-bonded SRAM-PIM, where read-out bandwidth from DRAM becomes the new performance bottleneck.}

\begin{figure}[h]
  \centering
  \includegraphics[width=0.9\linewidth, trim=0cm 0.95cm 0cm 0.3cm, clip]{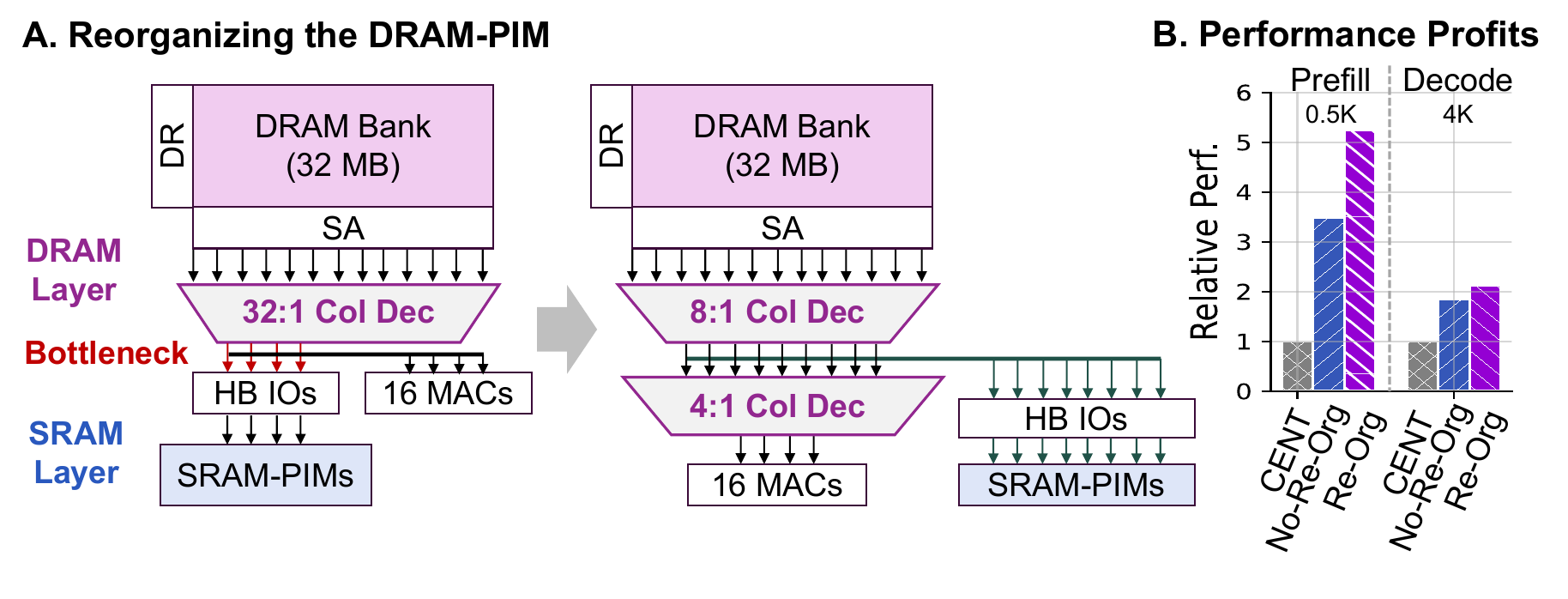}
  \caption{DRAM-PIM reorganization for CompAir can gain more performance profits taking Llama2-13B as the example.}
  \label{fig:reorg}
\end{figure}



To address this, we decouple the 32:1 column decoder to an 8:1 decoder for SRAM and a 4:1 decoder, substantially increasing bandwidth (Fig. \ref{fig:reorg}A). Applied to Llama-13B inference, this DRAM reorganization yields a 1.15–1.5$\times$ end-to-end speedup (Fig. \ref{fig:reorg}B). While this incurs a trade-off in I/O complexity or bond density, current hybrid bonding technologies (>10K/$mm^2$\cite{HybridBondingReview,ISSCC22HB}) support the extended bonds with just 10\% area of one DRAM bank, making this optimization both practical under current fabrication capabilities.

%% file: Sections/noc.tex
\section{In-Transit Computation with CompAir-NoC}
\label{sec:noc}

\subsection{Why CompAir-NoC}
\label{sec:whynoc}

We analyze the challenges of LLM non-linear computation in section \ref{ob_non_linear} and the need for efficient collective communication for matrix arithmetic performance in section \ref{sec:arch_sw_issue}.

\textcolor{black}{Actually, \textbf{data movement is unavoidable}, since device/channel/ bank-level parallelisms are necessary to fully utilize the resources of the scalable PIM, inevitably accompanied by data broadcasting and reduction. In addition, data movement exists between the PIM banks and NLUs. Therefore, what we need to do is to (i) minimize the additional data movement due to the non-linear computation and (ii) minimize the cost of the non-linear computation itself.}

\begin{figure}[h]
  \centering
  \includegraphics[width=0.95\linewidth, trim=0cm 0.1cm 0cm 0.2cm, clip]{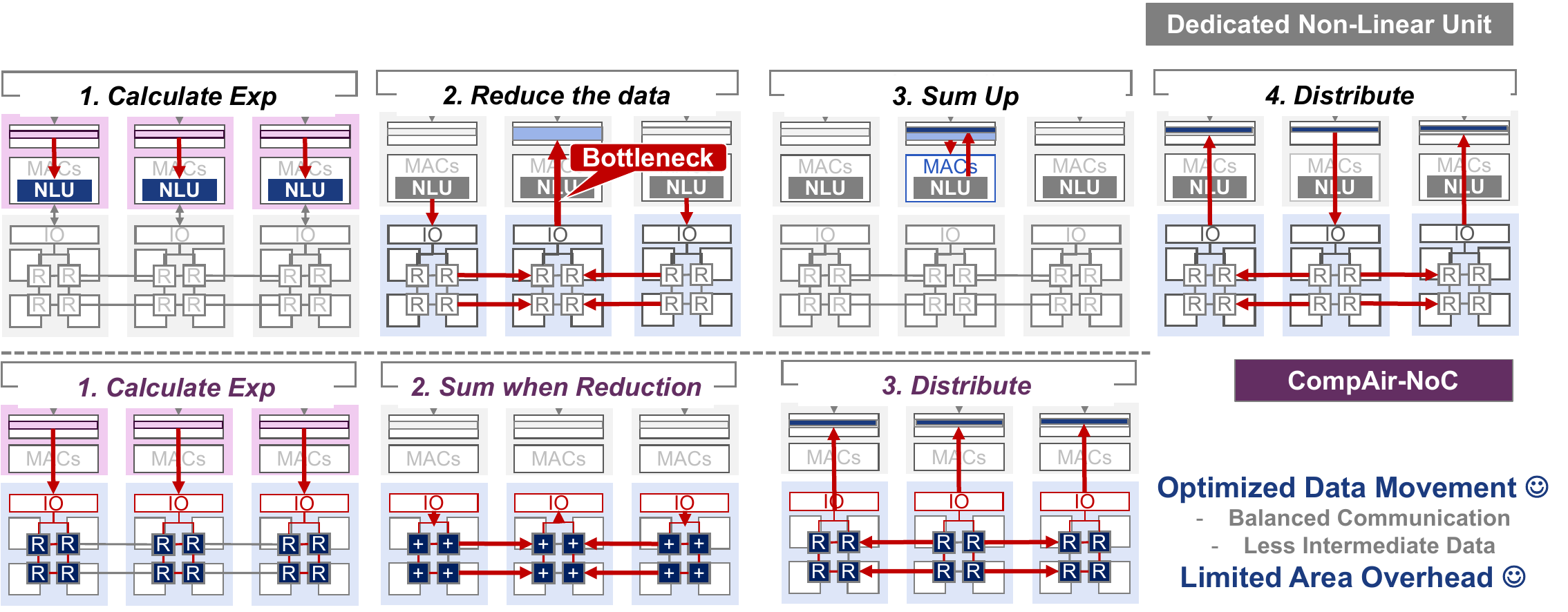}
  \caption{\textcolor{black}{The motivation of CompAir-NoC (Softmax)}}
  \label{fig:nocmotivation}
\end{figure}

\textcolor{black}{In section \ref{ob_non_linear}, we have revealed the non-negligible data movement overhead of centralized NLUs. Here we consider the method of the distributed NLUs, implementing an NLU for each DRAM-bank and taking Softmax as an example in Fig. \ref{fig:nocmotivation}. Each bank needs to use the NLU to perform the exponential computation. Then, the results of all the bank are summed up and distributed to every one. We find that (i) NLU is costly but idle in most of the time, (ii) summing and reduction are logically coupled, but physically completed by different devices, bringing data movement bottleneck. These inspire us to design a mechanism that can complete the computation when data moving, rather than generating redundant data movement to complete a certain operation, which brings benefits in two aspects:}

\textcolor{black}{\textbf{(i) Optimized Data Movement:} Computing during communication avoids the intermediate results from being moved (such as reduction) and prevents data from moving centrally towards a particular component, leading to congestion bottlenecks.}


\textcolor{black}{\textbf{(ii) Less Area Overhead:} If we can design a scheme that enables the arithmetic units multiplexing and streaming computation during communication, logic and buffer costs can be both saved compared to dedicated NLU.}

\textcolor{black}{For the opportunities that can be optimized in LLM, redundant data movements include three categories:} \textcolor{black}{\textbf{(i) Granularity Mismatching:} In RoPE, The swap of neighbouring scalars makes it necessary for a vector-based PIMs to perform scalar operations with NLUs or CPUs\cite{gu2025cent}.} \textcolor{black}{\textbf{(ii) Special Function:} PIM is difficult to perform some non-matrix multiplication operations, such as RMSNorm, SiLU, Softmax.} \textcolor{black}{\textbf{(iii) Collective Communication:} In Softmax and FC parallelism, reduction/broadcast brings inefficient data movement, which can be optimized by tree-based hardware.}





\textcolor{black}{Our goal is efficient non-linear computation and collective communication with minimal area and latency. The three challenges share: (1) fine-grained data movements, (2) tight computation-memory-communication coupling. NoCs serialize vectors into flits \cite{Jerger2017Book}, enabling fine-grained manipulation in CompAir. NoCs also support dynamic dataflow\cite{TPU, OpenCGRA}. We present CompAir-NoCs, a lightweight, computation-enabled NoC with inter-bank reconfigurability. Unlike prior work\cite{Sangaiah2020SnackNoC, Huang2019ActiveRouting, OmniComp}, CompAir-NoC supports diverse distributed computations without compromising communication. Section \ref{sec:router_micro_arch} details its microarchitecture, while Section \ref{noc4nlop} shows its efficiency for LLM non-linear operations.}

\subsection{CompAir-NoC Router Microarchitecture}
\label{sec:router_micro_arch}

\begin{figure}[h]
  \centering
  \includegraphics[width=0.95\linewidth, trim=0cm 0.3cm 0cm 0.4cm, clip]{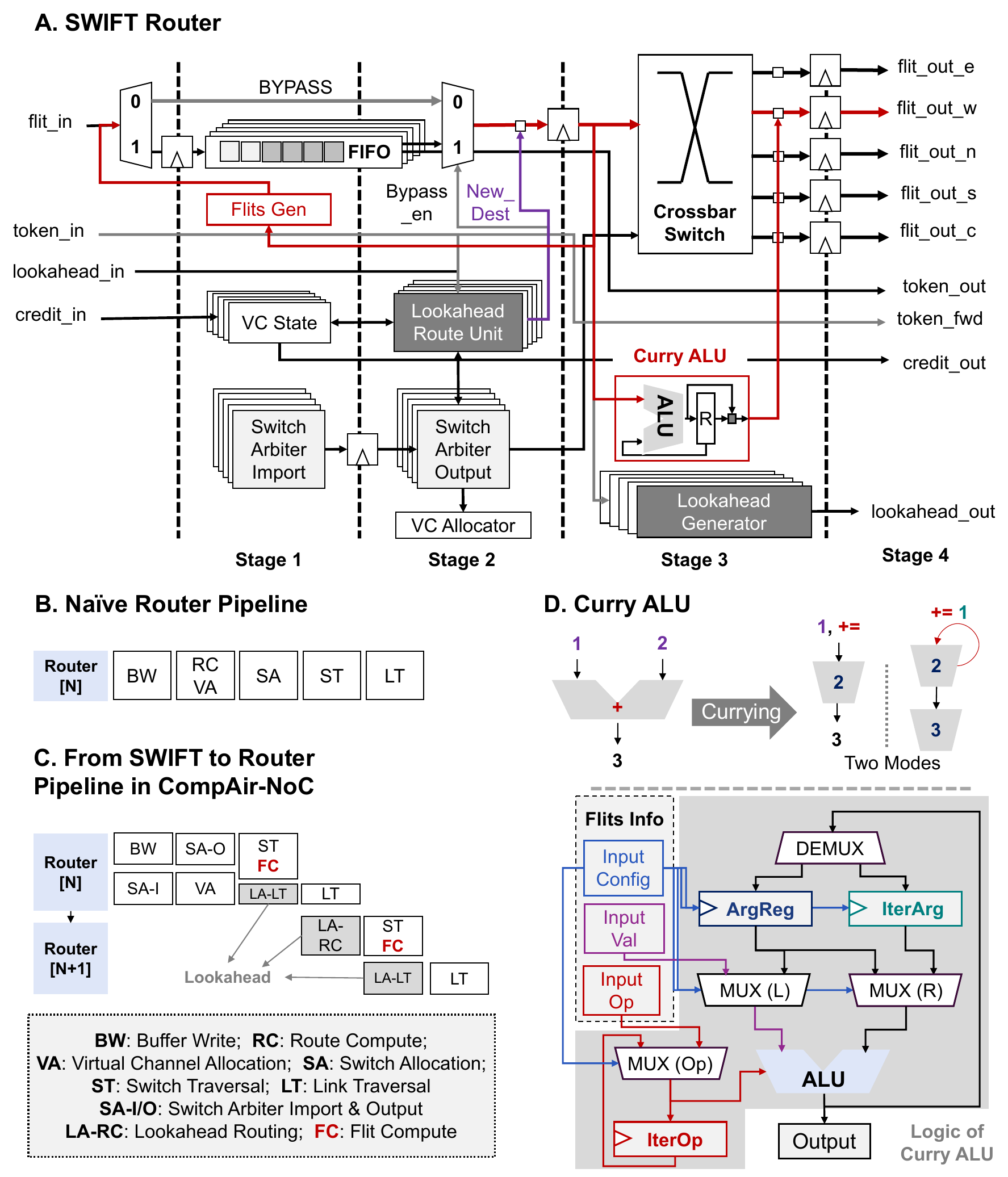}
  \caption{\textcolor{black}{CompAir-NoC router microarchitecture.}}
  \label{fig:router}
\end{figure}

Fig. \ref{fig:router}A (excluding red-highlighted parts) illustrates a classical optimized NoC architecture, SWIFT\cite{SWIFT1, SWIFT2}, where data is relayed in flits (32-128 bits) passing through the routers hop by hop. Unlike the simplest five-stage pipelined router (Fig. \ref{fig:router}B), the SWIFT router can compress the delay of a flit within a router to only 1-2 cycles with lookahead and bypassing (Fig. \ref{fig:router}C). This also means that any added computation must operate under light cycle budgets.


\textcolor{black}{Traditional dataflow requires dynamic operand matching across input flits, incurring significant latency and hardware overhead}\cite{Trips,WaveScalar}. Ideally, each flit can trigger the operation independently without waiting for others. Inspired by Currying in Lambda Calculus\cite{Hindley2008book}, we design an ALU driven by a single oprand, dubbed as \textbf{Curry ALU}. 

\vspace{-0.1cm}

\textcolor{black}{Currying transforms multi-operand functions into chained unary functions, where each function fixes some of the arguments and returns a new function awaiting the next.} Taking \verb|a*b+c| as an example, \verb|MulAdd(a,b,c)| is a multivariate function implementation, and the currying technique can equivalently convert it to: \verb|((curriedMulAdd(a))(b))(c)| where one input is accepted at a time, and taking ``function + partial oprands'' as a new function.

Fig. \ref{fig:router}D illustrates the fundamental idea in Curry ALU: most dataflow architecture dynamically transfer data, with operators statically bounded in the ALU\cite{SPU, OpenCGRA}; whereas Curry ALUs dynamically transfer a Currying function (a unary operator \verb|InputOp| and its left value \verb|InputVal|), with its internal \verb|ArgReg| statically storing the function parameters of the function (unary operator's right value). Curry ALU also contains the internal configurable \verb|IterArg| and \verb|IterOp| to allow \verb|ArgReg|'s iterated updating. Taking \verb|+=| as the example, an \verb|InputOp|-based mode would be \verb|InputVals+=ArgReg| (\textcolor{black}{Fig. \ref{fig:router}D left,} \verb|ArgReg| \textcolor{black}{is 2}), while an \verb|IterOp|-based mode would be \verb|ArgReg+=IterArg| (\textcolor{black}{Fig. \ref{fig:router}D right, }\verb|ArgReg| \textcolor{black}{is updated to 3}).

Curry ALU avoids flits matching and enables efficient \verb|ArgReg|-reuse. \textcolor{black}{Moreover, Curry ALU introduces minimal disruption to the high performance router pipeline.} The logical modifications caused by the Curry ALU are highlighted in red in Fig. \ref{fig:router}A. In Fig. \ref{fig:router}C, we use ``flit compute'' to mark the computation stage, which is parallel to the switch traversal. \textcolor{black}{In the flit compute stage, Curry ALU replaces the data in the original flit with the computation result in situ for overhead-free.}


\subsection{Supporting Non-Linear Operations in LLM}
\label{noc4nlop}

Then, we show how to implement flexible and localized non-linear operations in LLM based on CompAir-NoC.

\subsubsection{Data Rearrangement}



\begin{figure}[h]
  \centering
  \includegraphics[width=0.9\linewidth, trim=0cm 0.3cm 0cm 0.4cm, clip]{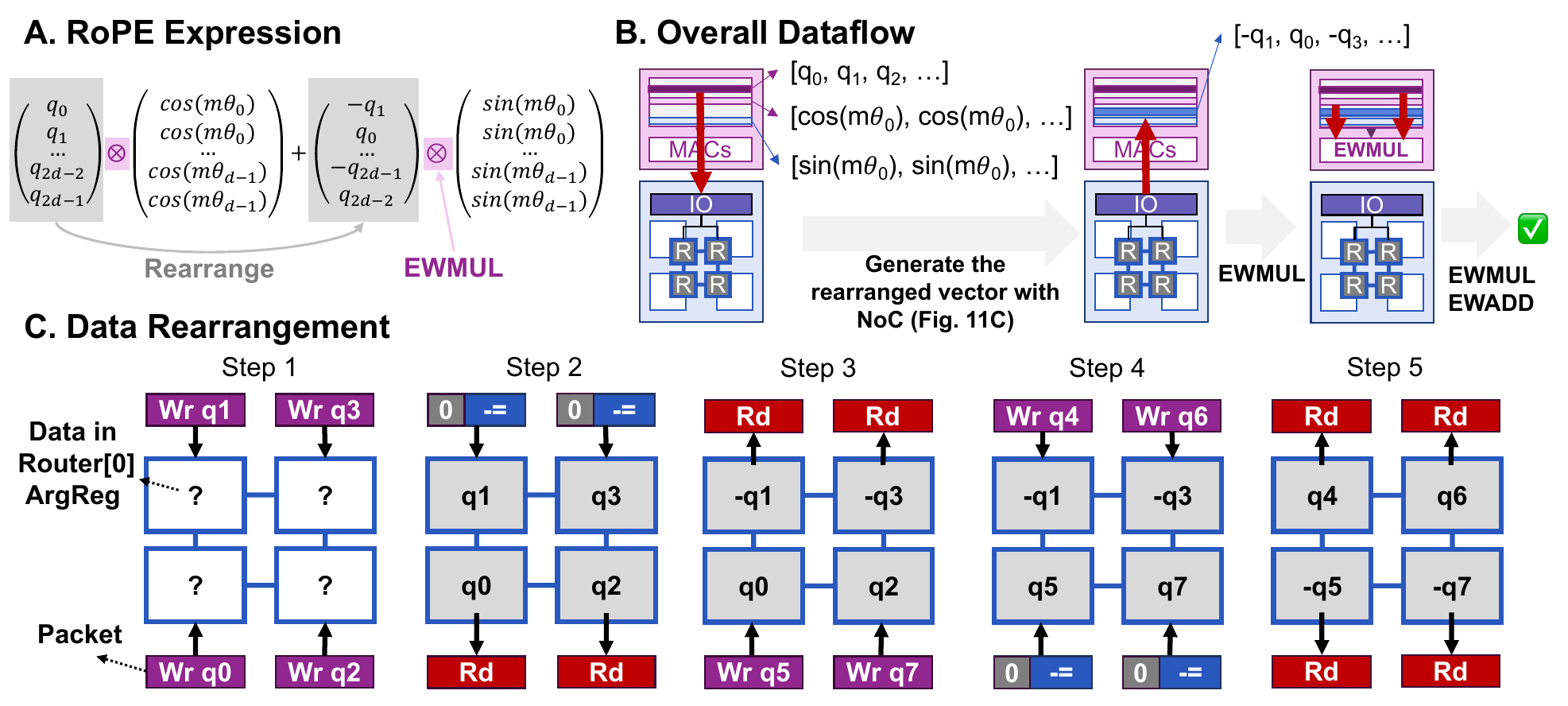}
  \caption{\textcolor{black}{RoPE data rearrangement with CompAir-NoC.}} 
  \label{fig:rope}
\end{figure}

\textcolor{black}{\textcolor{black}{To process RoPE in Fig. \ref{fig:rope}A, DRAM-PIMs (operating at row granularity) incur high overhead from frequent data transfers—shuttling data between DRAM banks and the CXL controller’s CPU for neighbor swaps and odd-digit negations.}}

The router provides the opportunity for fine-grained manipulation for RoPE, leveraging the \verb|ArgReg|s as the flexible buffer, \textcolor{black}{then letting DRAM-PIM implement efficient element-wise multiplication (EWMUL) as shown in Fig. \ref{fig:rope}B.} Fig. \ref{fig:rope}C shows that four routers in each bank can be utilized to achieve efficient data exchange by sending data in five stages. \textcolor{black}{In practical evaluation on the Llama2-7B, our approach completes the rearrangement of Q or K vectors in only 34 cycles per bank, with all banks executing in parallel, highlighting both scalability and speed.}


\subsubsection{Exponents and Square Root} Non-linear functions like exponents and square roots are central to Sigmoid and Softmax. In digital circuits, they are solved by iterative methods. The exponent and square root can be solved with Taylor expansion and Newton iteration, respectively.



\begin{figure}[h]
  \centering
  \includegraphics[width=0.95\linewidth, trim=0cm 0.1cm 0cm 0.4cm, clip]{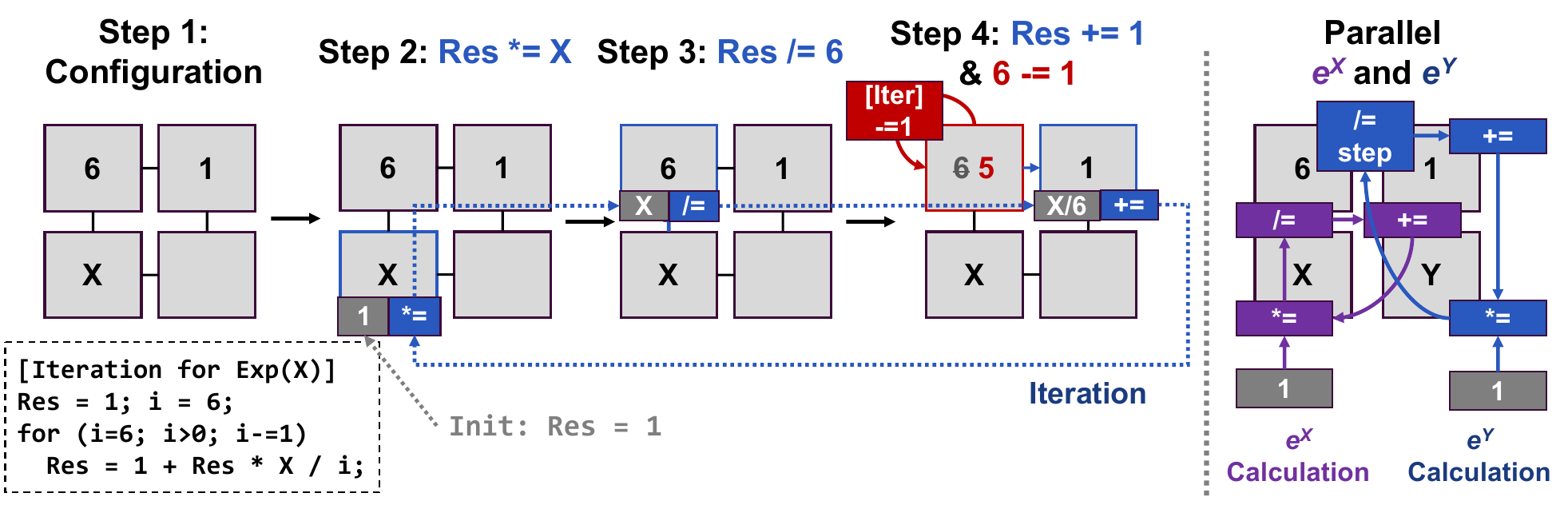}
  \caption{\textcolor{black}{Exponential function with CompAir-NoC.}}
  \label{fig:exp}
\end{figure}

Fig. \ref{fig:exp} presents an iterative computation method for the exponential function with dynamic \verb|ArgReg| updates. We configure the router with \verb|ArgReg=6| as iteration rounds (\verb|IterRound|), initialized with \verb|IterArg=1| and update operation \verb|IterOp='-='|. The computation proceeds outward from innermost levels, applying operations \verb|*=X|, \verb|/=IterRound|, and \verb|+=1| in each iteration until \verb|IterRound=0|. Our design enables efficient hardware utilization, supporting two parallel exponentiation across four routers. \textcolor{black}{In each channel, 16 banks enables 32 concurrent exponential functions in total}. This approach extends naturally to square root implementations.

\subsubsection{Broadcast Tree and Reduce Tree}

From the communication point of view, broadcast and reduce are inverse operations of each other from the tree structure. Taking reduction with a width of 16 as an example, it is equivalent to the existence of an operation function as: \verb|Reduction(‘+’,x[0],...,x[15])|. Therefore, it can be transformed into a 4-layer binary tree for parallel reduction, and we will use \verb|ArgReg| as the result of reduction for each non-leaf node to reduce. because the reduction of $2^N$ nodes theoretically requires $2^{N-1}$+$2^{N-2}$+... +1=$2^N-1$ intermediate nodes, so it can ensure that each node is fully utilized. In CompAir, we set the bank as the granularity for reduction, opening up more possibilities for linear operation improvements in DRAM-PIMs.

%% file: Sections/programming.tex
\section{Programming Model and ISA Design}
\label{sec:prog}

SIMD is the appropriate programming model for DRAM-PIM. However, SRAM-PIMs and router-level execution in CompAir-NoC introduce fine-grained operations and router-wise packet generation, demanding a MIMD appraoch. \textcolor{black}{This divergence creates \textbf{a conflict between SIMD and MIMD in terms of flexibility and programmability.} Two design choices exist here: \textit{(i)} unify the MIMD programming modes of SRAM-PIM and NoC under a DRAM-compatible SIMD regime; or \textit{(ii)} extend MIMD into DRAM-PIM.}

While prior architectures\cite{HBA} pursue the second way, integrating distributed controllers in each DRAM-Bank for autonomous MIMD execution, this approach incurs a 17\% area overhead and fails to scale efficiently with massive computing units. In contrast, CompAir adopts the first and more scalable route: reconciling MIMD flexibility with DRAM’s SIMD constraints with lower control complexity by packet encoding and autonomous path generation. 





To achieve this objective, we address by \textbf{setting up a hierarchical ISA.} The \textbf{Row-Level ISA} is programmed at the DRAM bank granularity in SIMD, and the \textbf{Packet-Level ISA} is granulated at the execution behaviour of the router. Moreover, the transformation from row-level instruction to packet-level instruction can be established directly. The row-Level ISA is a programming interface exposed to the user, while the packet-Level ISA is what the NoC-related instructions actually store in the instruction buffer after compilation. Then, sections \ref{sec:isa} are presented for each of these two levels. Section \ref{sec:trans} gives their translation rules.

\subsection{Row-Level ISA and Packet-Level ISA}
\label{sec:isa}

To program SRAM-PIM and NoC at row level in SIMD, the defined instructions are shown in Table \ref{tab:row_isa}. The \verb|SRAM_Write| and \verb|SRAM_Compute| instructions are used to write the weights for configuration and write input vector to the SRAM for computing. Each bank loads data from the same address (\verb|SRC|) and writes to the same address (\verb|DST|). The length is configured by \verb|Length| in the instruction. \textcolor{black}{CompAir's addressing is confined to DRAM banks, while SRAM-PIM operations (weight reloading and computing) are instruction-granular with fixed dataflow, eliminating SRAM addressing overhead.}

\begin{table}[h]
    \centering
    \small
    \setlength{\abovecaptionskip}{2pt}
    \caption{Row-Level ISA for NoC and SRAM-PIM}
    \begin{tabular}{ccccccc}
        \hline
        INST&OP&SRC&DST&NUM1&NUM2 \\ 
        \hline
        NoC\_Scalar&+=,-=,*=,/=&Addr&Addr&Mask&Config \\ 
        NoC\_Access&Rd, Wr&Addr&Addr&Mask&Const \\ 
        NoC\_BCast&/&Addr&Addr&Mask&SrcBank \\ 
        NoC\_Reduce&+=,-=,*=,/=&Addr&Addr&Mask&DstBank \\
        NoC\_Exchange&T+,T-,R+,R-&Addr&Addr&Offset&Group \\
        \hline
        SRAM\_Write&/&Addr&/&Length&/ \\ 
        SRAM\_Compute&/&Addr&Addr&Length&/ \\ 
        \hline
    \end{tabular}
    \label{tab:row_isa}
\end{table}

Within each bank, NoC-related instructions are operated at scalar granularity. From the programming perspective, we view the NoC purely as a computational component in this ISA level, without considering the communication behavior within the NoC. Five NoC-related instructions are designed.

\verb|NoC_Scalar| is responsible for once computation in router and \verb|NoC_Access| is used to read/write the Curry ALU's registers: the 64-bit \verb|Mask| is used to indicate whether 64 routers of a channel accept the computation task, and \verb|Const| is used to express the argument. 

\verb|NoC_Reduce| and \verb|NoC_BCast| denote reduction and broadcasting in DRAM bank granularity respectively. \verb|Mask| remains the same as \verb|NoC_Scalar|, indicating whether each SRAM-PIM \textcolor{black}{macro} participates in reductions and multicasts. Reductions and multicasts support parallel execution of 4 trees, so DstBank/SrcBank indicates the target bank for reductions or the source bank for multicasts.

\verb|NoC_Exchange| is quite different from the above design, as there may be intra-row as well as inter-bank data exchange. Therefore, in the coding of its operation, \verb|T| expresses that data is exchanged between the banks, while \verb|R| means data is exchanged within the rows of each bank. And + and - are used here to indicate whether inversion is required. The swapped two targets is specified with \verb|Offset| and \verb|Group|. The rule is that for position \verb|x|, it is exchanged with the position of (\verb|x|+\verb|Offset|)\%\verb|Group|. For RoPE, the exchange is expressed as \verb|NoC_Exchange(R-,SrcRow,DstRow,1,2)|.


\begin{table}[h]
    \centering
    \small
    \setlength{\abovecaptionskip}{2pt}
    \caption{Packet-Level ISA for NoC}
    \begin{tabular}{ccccccc}
    \hline
        Type&Data&IterNum&Path[0]&Path[1]&Path[2]&Path[3] \\ \hline
        4b&16b&4b&12b&12b&12b&12b \\ 
        \hline
    \end{tabular}
    \\[4pt]
    \begin{tabular}{cccccc}
    \hline
        Path.X&Path.Y&Path.WrReg&Path.IterTag&Path.Opcode \\ \hline
        4b&4b&1b&1b&2b \\ 
        \hline
    \end{tabular}
    \label{tab:pkt_isa}
\end{table}

Table \ref{tab:pkt_isa} shows the packet information at the time of router execution. Where \verb|Type| is used to indicate the instruction information, currently includes seven types: None, Scalar, Reduce, Exchange, Broadcast, Read, and Write. The \verb|Data| field contains BF16-formatted payload within the packet. \verb|IterNum| specifies the iteration count for the computational path, while \verb|Path| defines the router sequence for each computation step (supporting up to four relay nodes per loop). The control signals include: \verb|WrReg| for register write-enable in CurryALU, \verb|IterTag| which triggers dynamic \verb|ArgReg| updates via \verb|IterArg| and \verb|IterOp| after computation.

\subsection{Autonomous ISA Translation}
\label{sec:trans}

\begin{figure}[h]
  \centering
  \includegraphics[width=0.95\linewidth, trim=0cm 0.4cm 0cm 0cm, clip]{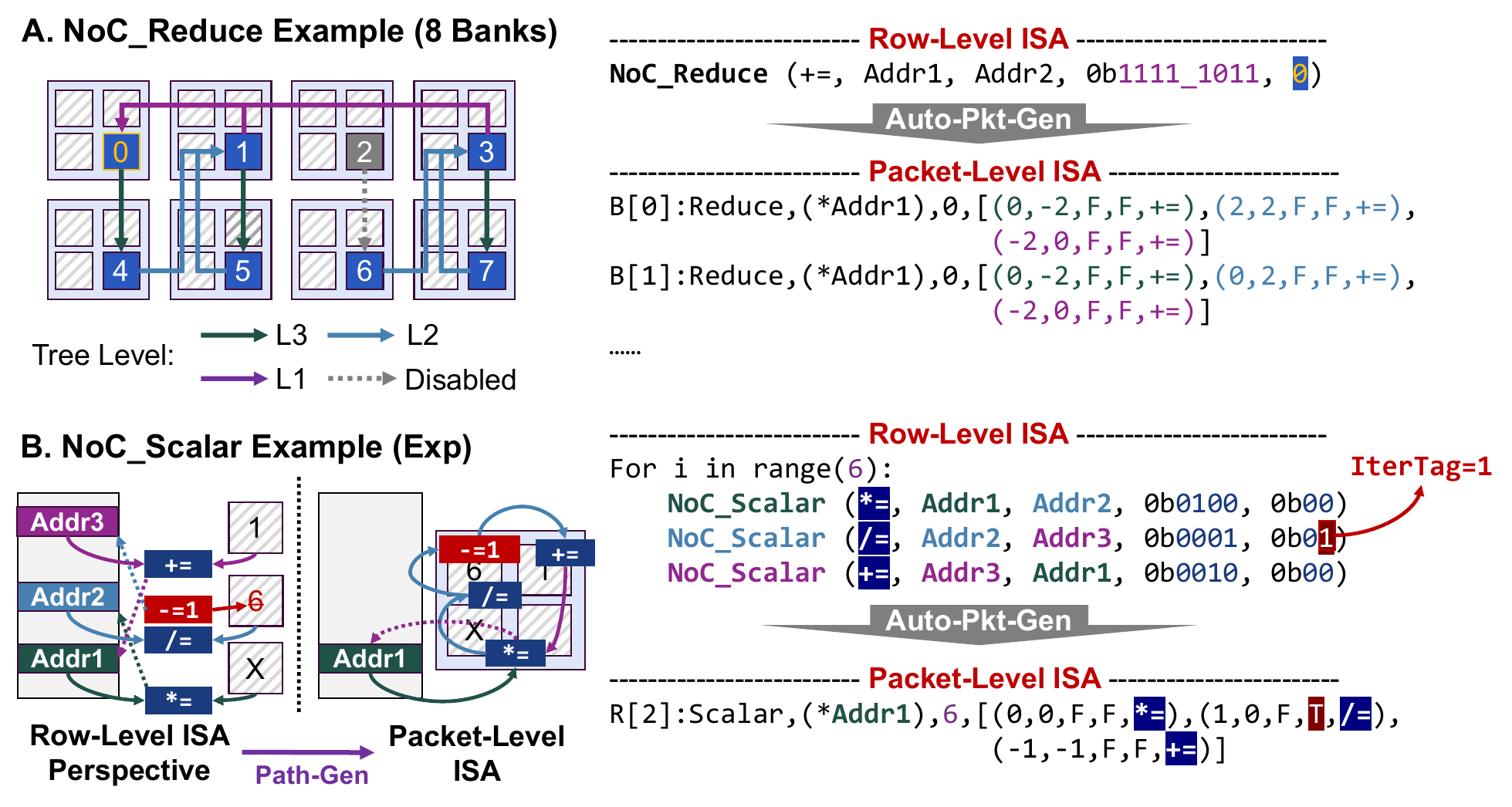}
  \caption{ISA translation. (A) NoC\_Reduce in 8 banks. (B) NoC\_Scalar for the iteration of exponential function.}
  \label{fig:isa}
\end{figure}

The two layers of ISA can be automatically converted. The key challenge in cross-level translation is that the row-level ISA fixes the data path of ``DRAM row $\rightarrow$ Curry ALU in router $\rightarrow$ DRAM row'' and ignores the behaviour of the NoC, which is exactly the part of the packet-level ISA that is needed. In Fig. \ref{fig:isa}, we show two typical transformation processes with \verb|NoC_Reduce| and \verb|NoC_Scalar|. 

\verb|NoC_Reduce| needs to instantiate an instruction into separate packets for each bank according to the bank id. Since the structure of the reduction tree is fixed, we design a dedicated pattern shown in Fig. \ref{fig:isa}A for automatic conversion between the two.

While our row-level ISA requires conservative write-back to DRAM for every \verb|NoC_Scalar| to maintain programming simplicity, this approach introduces significant inefficiencies and undermines the inherent flexibility of MIMD execution. Inspired by operation fusion techniques in AI compilers\cite{DNNfusion, TVMopfusion}, we propose a path generation mechanism that enables row-level ISA merging by fusing dependent \verb|NoC_Scalar| operations. Our key insight is that consecutive \verb|NoC_Scalar| operations can form a producer-consumer chain, where the \verb|DST| of one instruction serves as the \verb|SRC| of the next. By analyzing these sequences, we merge compatible operations into a single path in one packet, which encapsulates the entire computation and communication pattern. This optimization allows each router within a bank to execute the fused operation with just one packet, greatly simplifying the SRAM-PIM controller’s logic.

%% file: Sections/evalutionv2.tex
\section{Methodology}

The CompAir is implemented with cycle-accurate simulators. The DRAM and NoC are simulated with ramulator2.0\cite{Ramulator2} and Booksim\cite{booksim}. The SRAM-PIM is based on the chip specifications from \cite{SRAMpimISSCC23}. The inter-device communication and DRAM-PIM instruction execution are based on the CENT simulator\cite{gu2025cent}. To evaluate the area cost of CompAir-NoC, we implement the RTL of CompAir-NoC and synthesize the corresponding area report with Synopsys Design Compiler. The UMC 28nm process library is used for evaluation. For the choice of baseline, fully DRAM-PIM based CENT\cite{gu2025cent} and AttAcc\cite{AttAcc} with HBM-PIM and A100 hybrid architecture are chosen for comparison. We test them with a number of different LLM models at different sequence lengths, batch sizes, and parallelism strategies, including the Llama series (7B, 13B, 70B)\cite{llama2}, Qwen-72B\cite{qwen2}, and GPT3-175B\cite{gpt3}. \textcolor{black}{The hardware configuration of CompAir is shown in Table \ref{tab:hardware_config}.}


\vspace{-0.1cm}
\begin{table}[htbp]
\centering
\setlength{\abovecaptionskip}{2pt}
\caption{\textcolor{black}{Hardware Configurations for Evaluation}}
\label{tab:hardware_config}
\begin{tabular}{cl}
\toprule
\textbf{Component} & \textbf{Specification} \\
\toprule
& 32 channels/device, 16 banks/channel, BF16 \\
DRAM-PIM& 32MB/bank, 16 MACs/bank, $t_{RCDWR}$=14ns \\
\cite{gu2025cent}& $t_{RCDRD}$=18ns, $t_{RAS}$=27ns, $t_{CL}$=25ns, $t_{RP}$=16ns \\
\cmidrule(lr){1-2}
SRAM-PIM & 64kb for each array, BF16, 4 arrays/bank \\
\cite{SRAMpimISSCC23}& $t_{access}$ = 6.8-14.1ns, 14.4-31.6TOPS/W (0.9-0.6V) \\
\cmidrule(lr){1-2}
& 4$\times$16 2D-mesh, 2 BF16 Curry ALUs per router \\
CompAir-NoC& 1 adder/multiplier/divider per ALU, flit size: 72b \\
& routing: DOR, router arch: SWIFT\cite{SWIFT1} \\
\bottomrule
\end{tabular}
\end{table}

\section{Evaluation}
\label{sec:evaluation}

\begin{figure}[h]
  \centering
  \includegraphics[width=0.9\linewidth, trim=0cm 0.4cm 0cm 0cm, clip]{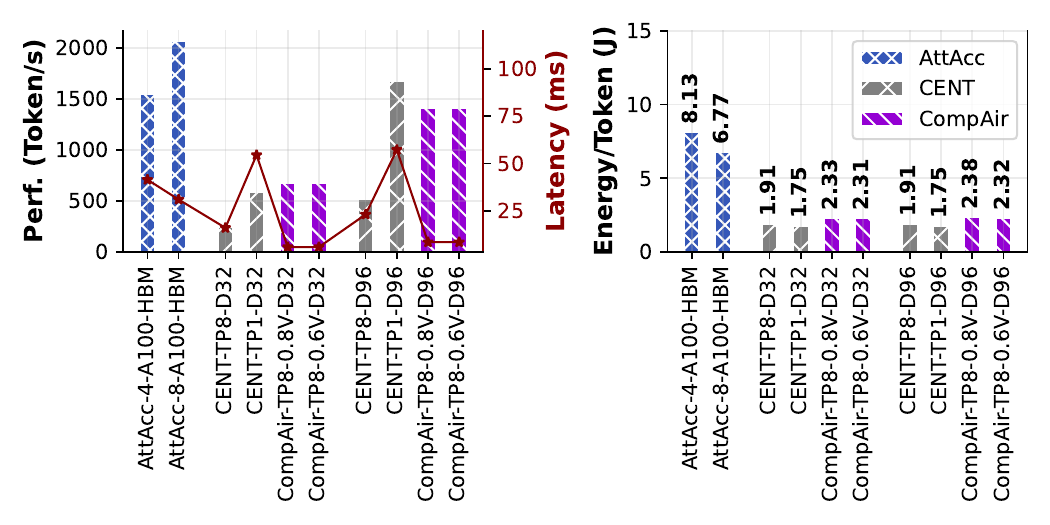}
  \caption{Energy per token and performance analysis \textcolor{black}{(Batch=64, Decode, Seqlen=128K)} between CompAir, CENT (GDDR6-PIM)\cite{gu2025cent}, and AttAcc (Nvidia A100 GPU + HBM-PIM)\cite{AttAcc} with GPT3-175B\cite{gpt3}. \textcolor{black}{``AttAcc-4-A100-HBM'' refers to 4 80GB A100 and 4 16GB HBM3-PIM devices are used.}}
  \label{fig:power}
\end{figure}

\textcolor{black}{Firstly, we will conclude the thinking behind the experiments carried out in this paper. In the previous sections, our experiments have demonstrated that (1) pure SRAM-PIM is unrealistic for LLM (Fig. \ref{fig:sram_vs_dram}A). (2) SRAM-PIM and DRAM-PIM have advantages in batched FC and Attention, respectively (Fig. \ref{fig:sram_vs_dram}B, \ref{fig:sram_vs_dram}C, \ref{fig:nocmotivation}, \ref{fig:gqa_latency}), and making it valuable to hybridize the two. (3) CompAir-NoC can eliminate data movement from centralized NLUs (Fig. \ref{fig:acc}). The questions we need to further validate in this section are (1) how much performance improvement (Fig. \ref{fig:power},\ref {fig:llama270b}, \ref{fig:gqa_latency}) and energy cost (Fig. \ref{fig:power}, \ref{fig:gqa_energy}) hybrid PIM can bring compared to pure DRAM-PIM. (2) The impact of different LLM configurations on this performance improvement and which design plays a role in it respectively (Fig. \ref{fig:prefill}-\ref{fig:longcontext}). (3) The hardware cost (Fig. \ref{fig:area}) and benefit (Fig. \ref{fig:nlr}, \ref{fig:pathgen}) of CompAir-NoC.}

\vspace{-0.3cm}

\subsection{End-to-End Performance}

\textbf{Firstly, we conduct an overall evaluation of CompAir's latency, throughput, and energy consumption.} The results are shown in Fig. \ref{fig:power}, where we evaluated CENT and CompAir according to the 32 device and 96 device cases, respectively. The full pipeline parallism (PP) approach is used in the original CENT and AttAcc comparison experiments\cite{gu2025cent}, but our experiments find that this causes a significant increase in the latency of individual tokens. Therefore, we choose a relatively balanced configuration of 8-device tensor parallelism (TP=8). The results show that CompAir achieves better throughput and latency than CENT for 32- and 96-device scaling in the same configuration. The \textcolor{black}{throughput} of 96 devices is comparable to the throughput of Attacc (4 A100s and 4 HBMs), but the latency and energy consumption per token are only 20.2\% and 28.5\% of AttAcc in 4K context. \textcolor{black}{In details, Fig. \ref{fig:power}A shows that CompAir achieves almost equal proportional latency and throughput performance gains compared to the equivalent parallel strategy of CENT (TP = 8). In Fig. \ref{fig:power}B, CompAir increases energy compared to pure DRAM-PIM due to cross-die communication. Optimizing the DRAM-PIM/SRAM-PIM ratio enables latency gains with modest energy overhead versus DRAM-PIM-only, but excessive use of SRAM-PIM risks high energy costs (further analyzed in Fig. \ref{fig:gqa_energy}).}

\begin{figure}[h]
  \centering
  \includegraphics[width=0.9\linewidth, trim=0cm 2cm 0cm 0.3cm, clip]{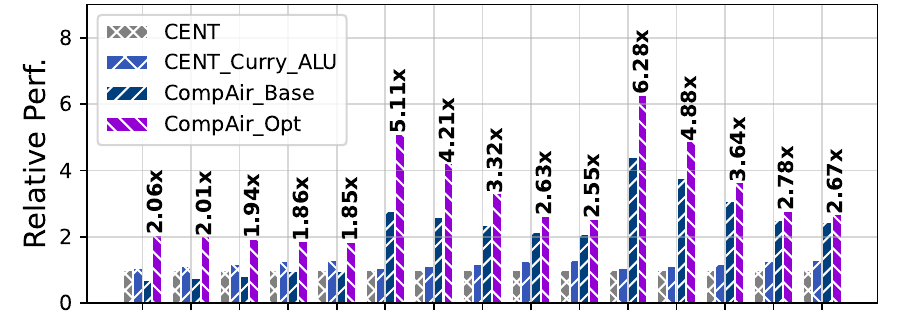}
  \includegraphics[width=0.9\linewidth, trim=0cm 0.2cm 0cm 0.35cm, clip]{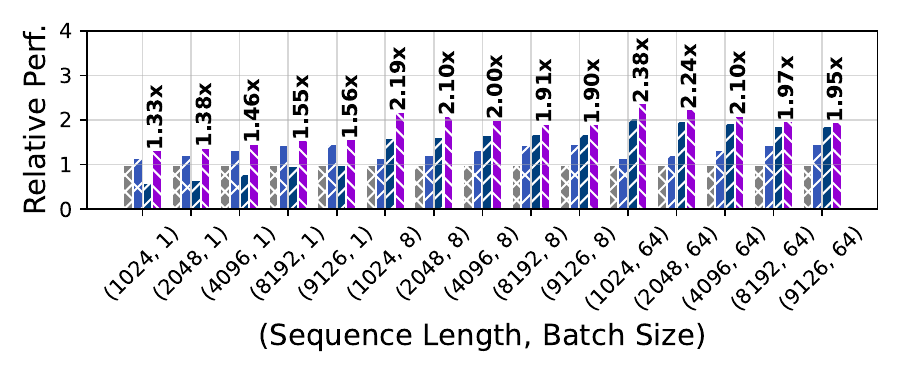}
  \caption{\textcolor{black}{Llama2-70B (Up) and Llama2-7B (Down) throughput evaluation with difference batch sizes and sequence length \textcolor{black}{for decode stage}.}}
  \label{fig:llama270b}
\end{figure}

\textbf{Next, we perform ablation experiments, sensitivity analysis and cost analysis of CompAir's performance gains.} For simplicity, we use CENT as the baseline and disassemble the performance as: \textit{(i)} \verb|CENT_Curry_ALU|: the full DRAM-PIM system combined with the localized Curry ALU. \textit{(ii)} \verb|CompAir_Base|: enabling SRAM-PIM but not modifing the DRAM-PIM's column decoder. \textit{(iii)} \verb|CompAir_Opt|: optimized CompAir with decoupled column decoder.

\begin{figure}[h]
  \centering
  \includegraphics[width=0.9\linewidth, trim=0cm 0.5cm 0cm 0cm, clip]{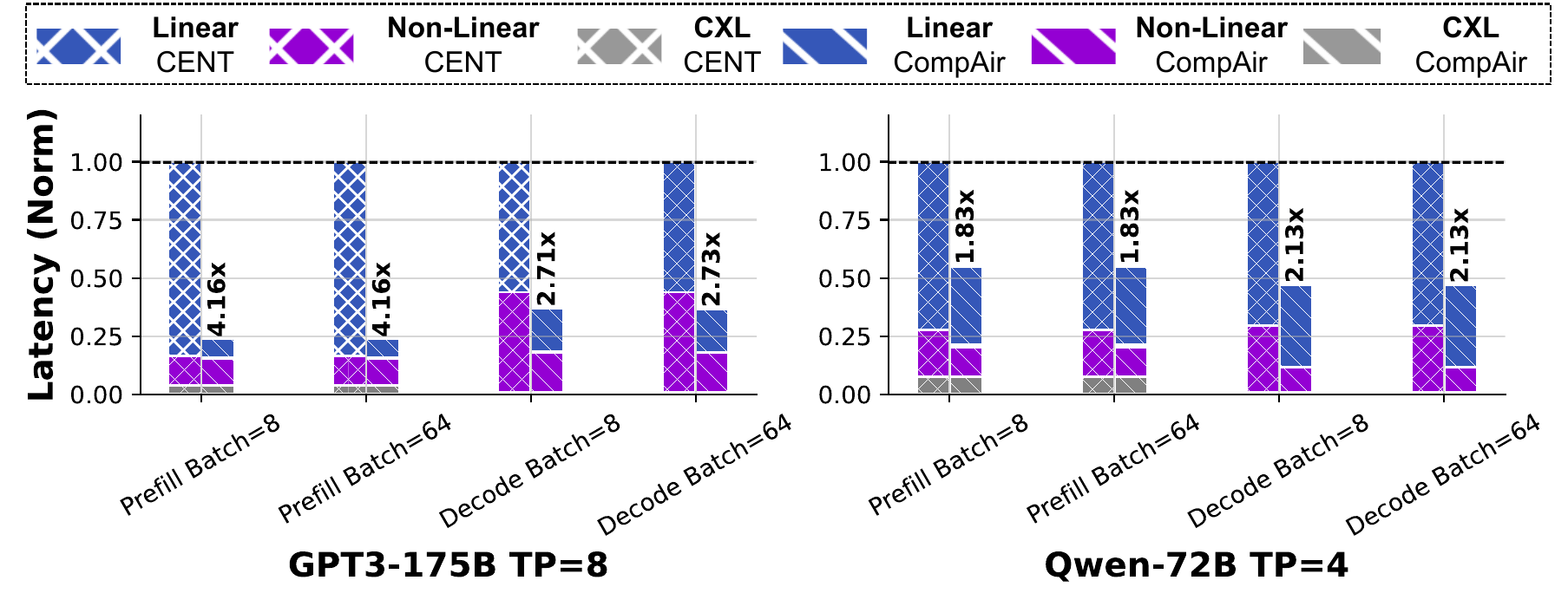}
  \caption{Long context evaluation with Qwen-72B\cite{deepseek, qwen2} and GPT3-175B\cite{gpt3} with 128K sequence length and 8K generation length (left bar: CENT, right bar: CompAir).}
  \label{fig:longcontext}
\end{figure}

In Fig. \ref{fig:llama270b}, the decode stages of Llama2-70B and Llama2-7B are used as an example to demonstrate the \textcolor{black}{throughput benifit} of CompAir under different sequence lengths and batch sizes. The results show that at batch size of 1, the introduction of SRAM-PIM does not bring better performance gain because the data reuse opportunity is limited. \textcolor{black}{When the batch size increases significantly, this advantage increases significantly and reaches a greater improvement of more than 2.67-6.28 times throughput in 64 batches}. As the sequence length increases, the relative \textcolor{black}{throughput} advantage stabilizes at approximately 2.5$\times$, indicating limited overall improvement. However, the contribution from the Curry ALU becomes more significant for longer sequence length. We will further analyze the performance in scenarios with very long context in Fig. \ref{fig:longcontext}.

\begin{figure}[h]
  \centering
  \includegraphics[width=0.95\linewidth, trim=0cm 0.3cm 0cm 0cm, clip]{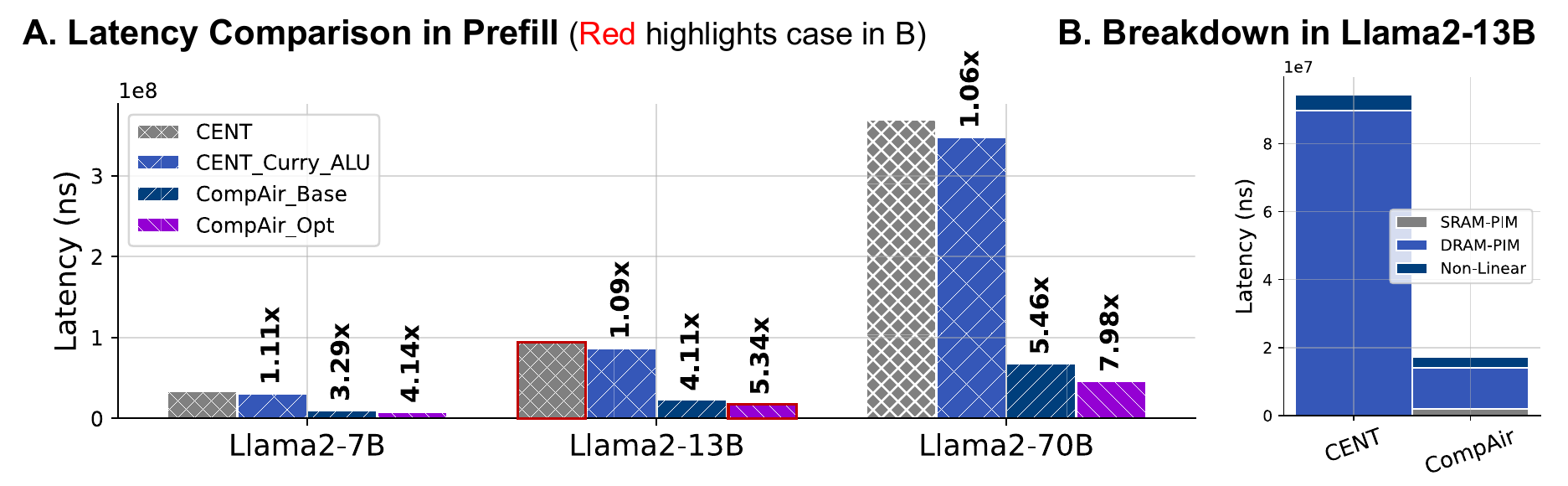}
  \caption{\textcolor{black}{Prefill stage with 0.5K generation length.}}
  \label{fig:prefill}
\end{figure}

Fig. \ref{fig:prefill} presents the performance of the compute-intensive prefill. With a 0.5K length, the SRAM-based PIM architecture achieves significant performance improvements ranging from 3.29$\times$ to 5.46$\times$ across various models. Furthermore, augmenting the DRAM read-out bandwidth yields additional performance gains, elevating the speedup ratio to between 4.1$\times$ and 7.89$\times$. \textcolor{black}{The performance gains of CompAir-NoC are limited in the short context, when the costs of data movement and non-linear computation are not bottlenecks.}

\begin{figure}[h]
  \centering
  \includegraphics[width=0.9\linewidth, trim=0cm 0.5cm 0cm 0.2cm, clip]{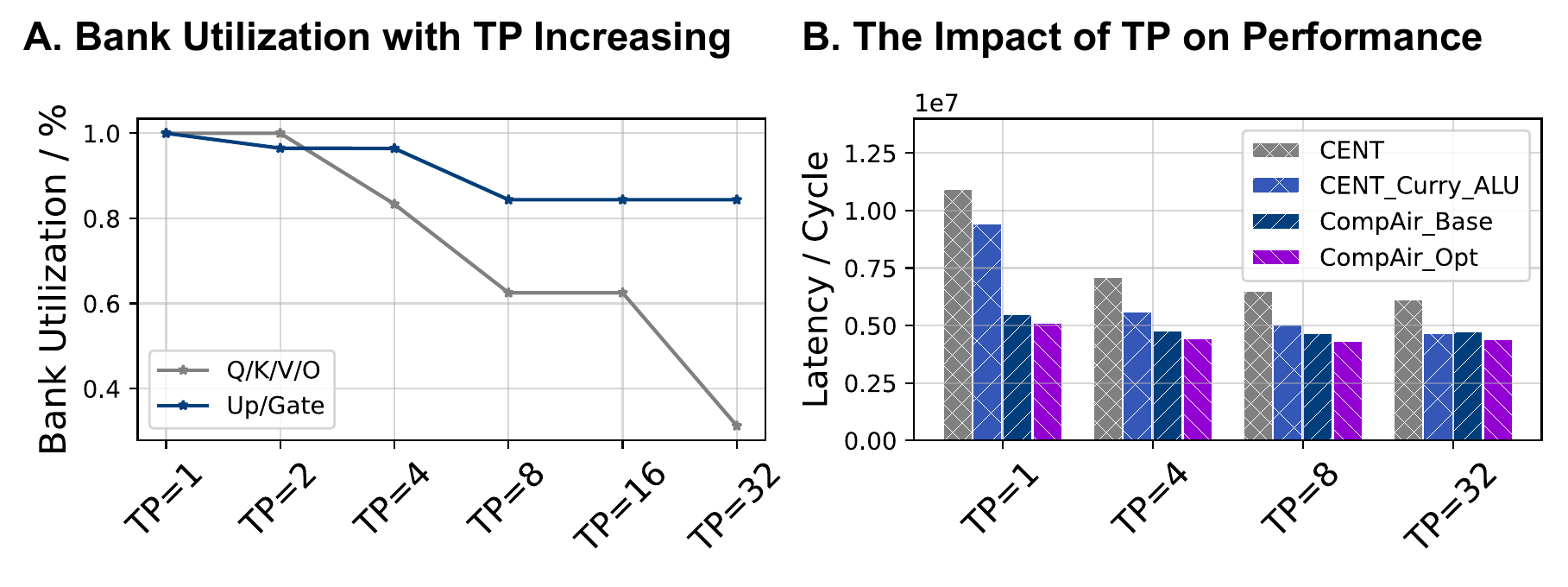}
  \caption{\textcolor{black}{Tensor parallelism evaluation with Llama2-13B. (A) The bank utilization drops rapidly if TP is too much. (B) The impact of TP on latency (Batch=64, Decode, Seqlen=4K).}}
  \label{fig:tp}
\end{figure}

To investigate the impact of parallelism strategies, we systematically evaluate various TP configurations from 1 to 32 devices. Our analysis reveals that both DRAM-PIM and CompAir exhibit latency convergence at high TP degrees due to substantially reduced bank utilization (Fig. \ref{fig:tp}). We have illustrated in Fig. \ref{fig:power} that larger TP configurations also incur significant throughput degradation. Consequently, we establish TP$\leq$8 as the optimal configuration range for most models. Within this range, CompAir maintains notable performance advantages, delivering 1.5-2.14$\times$ end-to-end speedup in Llama2-13B model. \textcolor{black}{Results show SRAM-PIM’s performance edge over DRAM-PIM stems from better data reuse. Increasing parallelism reduces this advantage by limiting reuse per bank, but also leads to an increase in data movement, when the latency reduction from CompAir-NoC becomes more significant.}

The above analysis allows us to draw a preliminary conclusion that SRAM-PIM can bring a very significant latency advantage for multi-batch scenarios, but the sequence length above are still within 10K. In Fig. \ref{fig:longcontext}, we test it for very long sequence scenarios with 128K decode and 8K prefill. For GPT3-175B and Qwen-72B, CompAir can bring 2.13-2.73 times performance improvement in the decode stage, thus illustrating the potential performance benefits of CompAir for the long sequence. Moreover, the proportion of nonlinear operation increases significantly, \textcolor{black}{revealing the benifits of CompAir-NoC when the context length increases}. CompAir-NoC reduces the non-linear latency, thus achieving performance improvement.

In all, \textcolor{black}{hybrid SRAM-PIM and DRAM-PIM in CompAir} exhibits significant improvement in prefill and multi-batch decode, while CompAir-NoC greatly optimizes long-context inference. \textit{CompAir offers considerable \textcolor{black}{latency} optimization for both MHA-bottleneck and FFN-bottleneck scenarios}.

\subsection{Micro-Architectural Evaluation}

The above experiments focus on the systematic performance evaluation under different workloads conditions. Then we focus on analyzing and evaluating the parameters for the microarchitecture.

\begin{figure}[h]
  \centering
  \includegraphics[width=\linewidth, trim=0cm 0.5cm 0cm 0.2cm, clip]{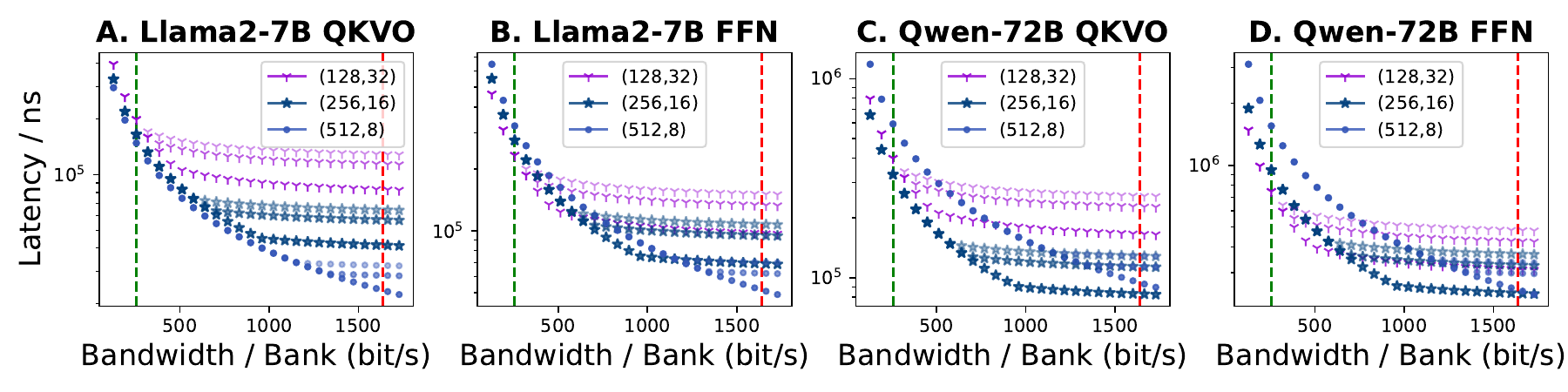}
  \caption{DSE of SRAM-PIM \textcolor{black}{in CompAir}. The lighter dots mark the \textcolor{black}{latency} at lower voltages (0.6V-0.8V).}
  \label{fig:dse}
\end{figure}

Fig. \ref{fig:dse} further provides a design space exploration (DSE) of SRAM-PIM in CompAir. In each subfigure, the green line marks the bandwidth in 32MB GDDR for each bank, and the red line marks the maximum bandwidth offered by hybrid bonding (6.4 Gbps). In this paper, we find that different macro configuration shapes produce a divergence point, before which different voltage configurations of SRAM-PIM do not affect the final performance since the \textcolor{black}{latency} is mainly affected by the input bandwidth. However, after the divergence point, the latency of the SRAM-PIM accounts for the main influence. The relative \textcolor{black}{latency} of different configuration shapes under different workloads is not fixed. Overall, wider inputs perform better in larger bandwidths.

\begin{figure}[h]
  \centering
  \includegraphics[width=0.9\linewidth, trim=0cm 0.5cm 0cm 0.3cm, clip]{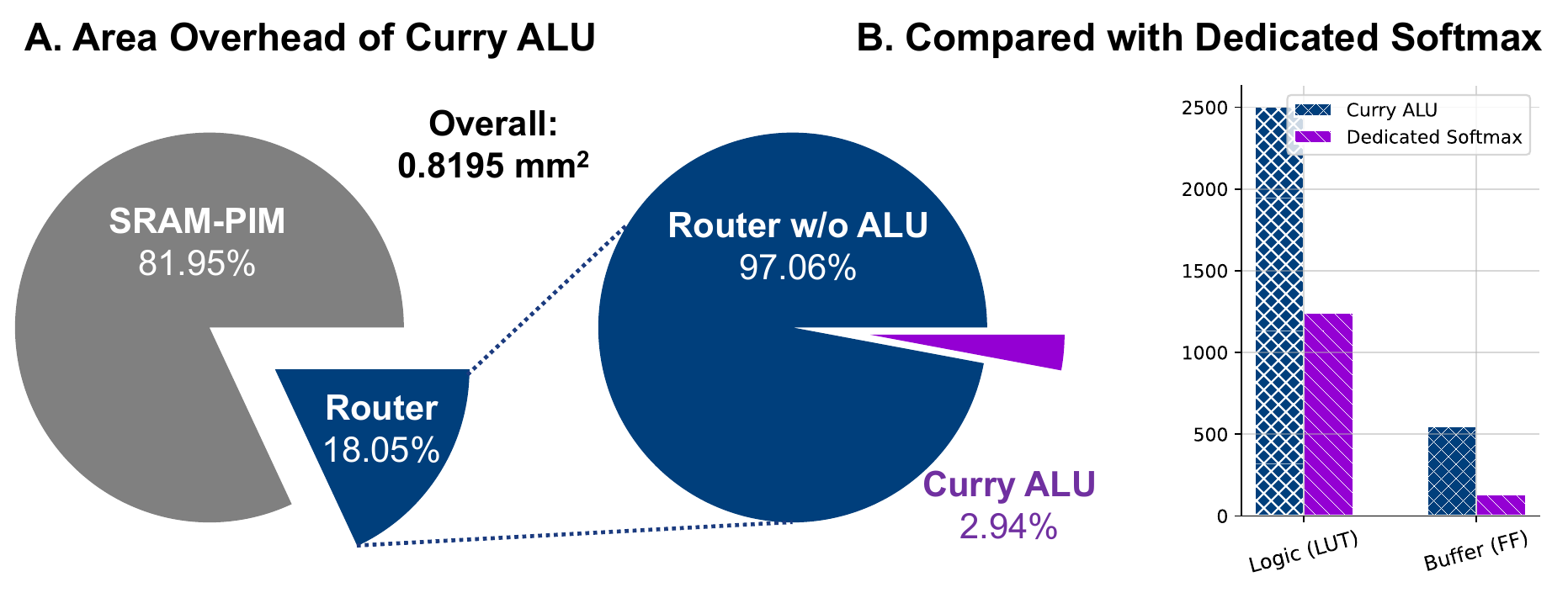}
  \caption{\textcolor{black}{Area overhead of Curry ALU.}}
  \label{fig:area}
\end{figure}

\begin{figure}[h]
  \centering
  \includegraphics[width=0.9\linewidth, trim=0cm 0.5cm 0cm 0.4cm, clip]{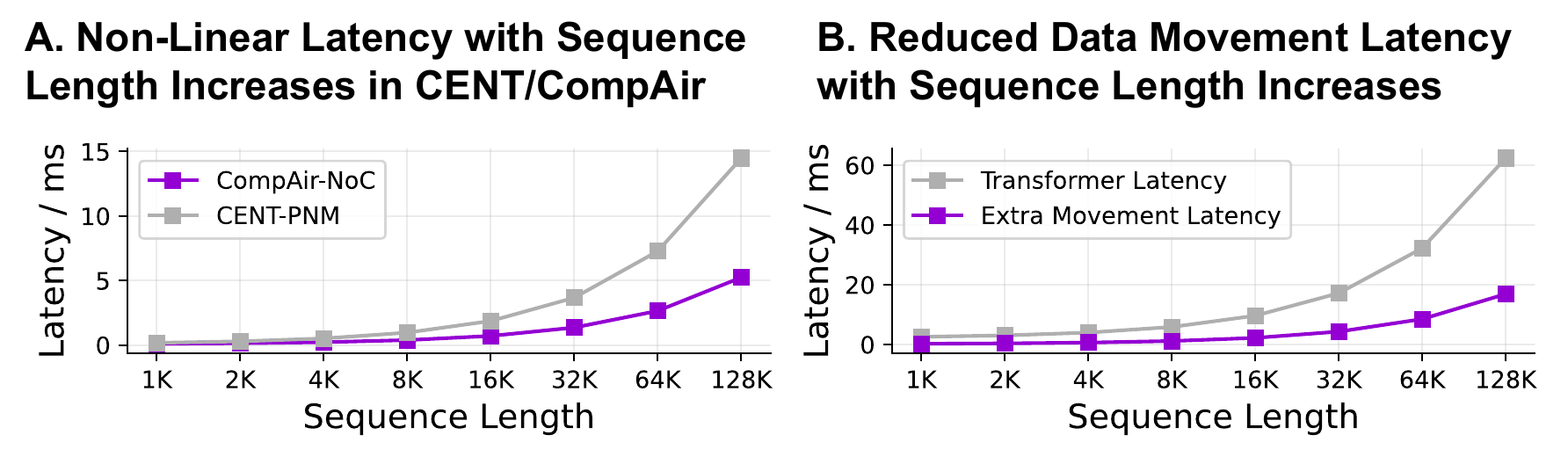}
  \caption{\textcolor{black}{Latency} profits from Curry ALU.}
  \label{fig:nlr}
\end{figure}

Then, we evaluate CompAir's area cost in Fig. \ref{fig:area}. The results show that the total area of SRAM-PIM and Router per bank is 0.8195$mm^2$, which fully satisfies the 3D stacking requirement of DRAM-PIM, and Curry ALU's area cost is only 2.94\% of router area. \textcolor{black}{We further compare the logic and memory resources used after synthesis of four Curry ALUs and one customised 16-input Softmax hardware unit using Vivado in Fig. \ref{fig:area}B. The results show that in the Curry ALUs use significantly less resources because computation in NoCs essentially perform stream processing to significantly reduce buffer usage.} The \textcolor{black}{latency profits} are also significant (Fig. \ref{fig:nlr}), as we specifically compare Curry ALUs to centralized non-linear computation units, compressing the total latency of non-linear computation by 30\% and optimizing the latency in long text by 25\%.

\begin{figure}[h]
  \centering
  \includegraphics[width=0.85\linewidth, trim=0cm 0.3cm 0cm 0.45cm, clip]{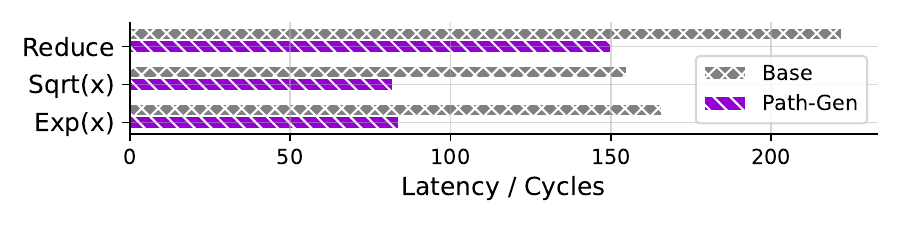}
  \caption{\textcolor{black}{Latency} profits from path generation.}
  \label{fig:pathgen}
\end{figure}

Finally, we evaluate the effectiveness of path generation in Fig. \ref{fig:pathgen}. \textcolor{black}{Base means that the data stream only supports SIMA style: IO buffer to Curry ALU and back to IO buffer.} Taking advantage of the NoC flexibility, a latency optimization of 33\%-50\% can be achieved compared to the row-level ISA without path generation. 

\vspace{-0.1cm}

\vspace{-0.3cm}

\section{Discussion}
\label{sec:discussion}

\textcolor{black}{In the evaluation, attention is processed by the DRAM-PIM, because matrix (K and V) are hardly reused between batches in both MLA\cite{deepseek} and MHA\cite{qwen2}. However, for GQA in LlaMa2-70B and Llama3\cite{llama2}, K and V are shared by a group of heads, enabling SRAM-PIM to accelerate attention. In this section, we analyze this in depth.}

\begin{figure}[h]
  \centering
  \includegraphics[width=0.9\linewidth, trim=0cm 0.3cm 0cm 0.4cm, clip]{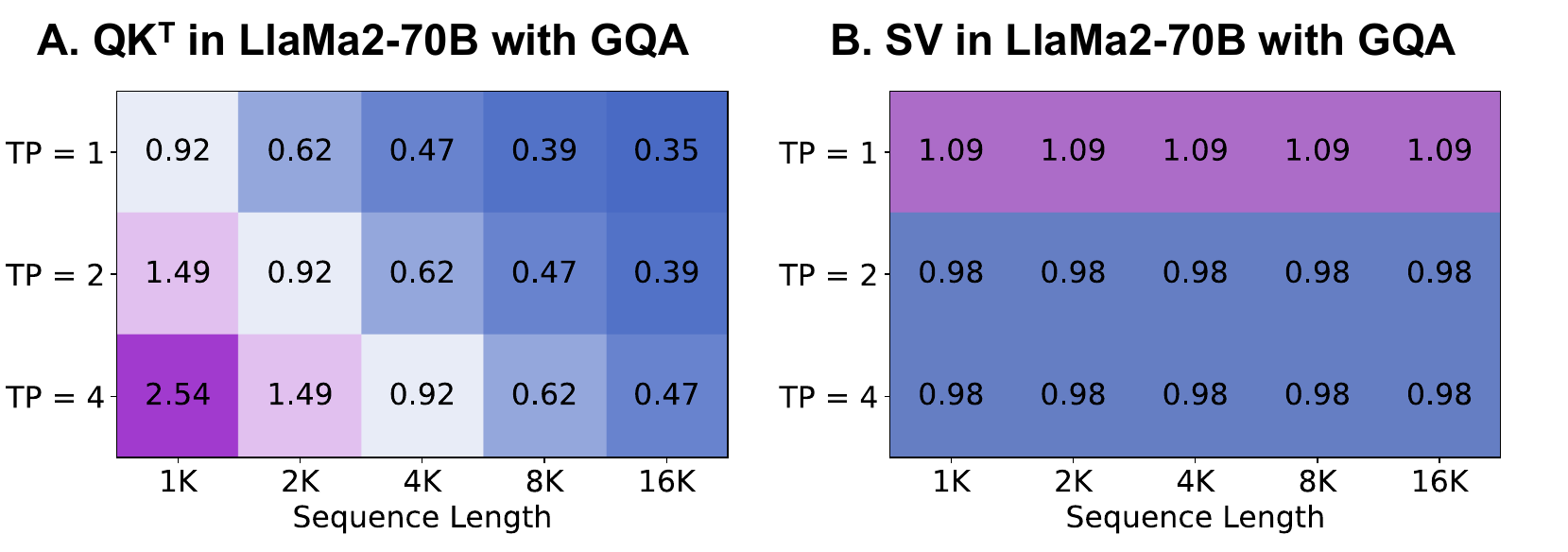}
  \caption{\textcolor{black}{The latency ratio between SRAM-PIM stacking DRAM and pure DRAM-PIM.} \textcolor{violet}{Purple}/\textcolor{blue}{blue} represents DRAM-PIM/SRAM-PIM stacking DRAM is better respectively.}
  \label{fig:gqa_latency}
\end{figure}

\textcolor{black}{Fig. \ref{fig:gqa_latency} evaluates the relative latency of DRAM-PIM and SRAM-PIM stacking DRAM with different sequence length and TP policies. In both, TP plays the role of splitting the $K^T$ and $V$ matrices to each bank along the dimension of the sequence length. For the SRAM-PIM, we correspond the sequence length dimension to the batch and the output dimension to the group size (8 in LlaMa2-70B), where the input dimension is hidden size primed by group size in $QK^T$ and equal to sequence length in $SV$.}

\begin{figure}[h]
  \centering
  \includegraphics[width=0.95\linewidth, trim=0cm 0.3cm 0cm 0.2cm, clip]{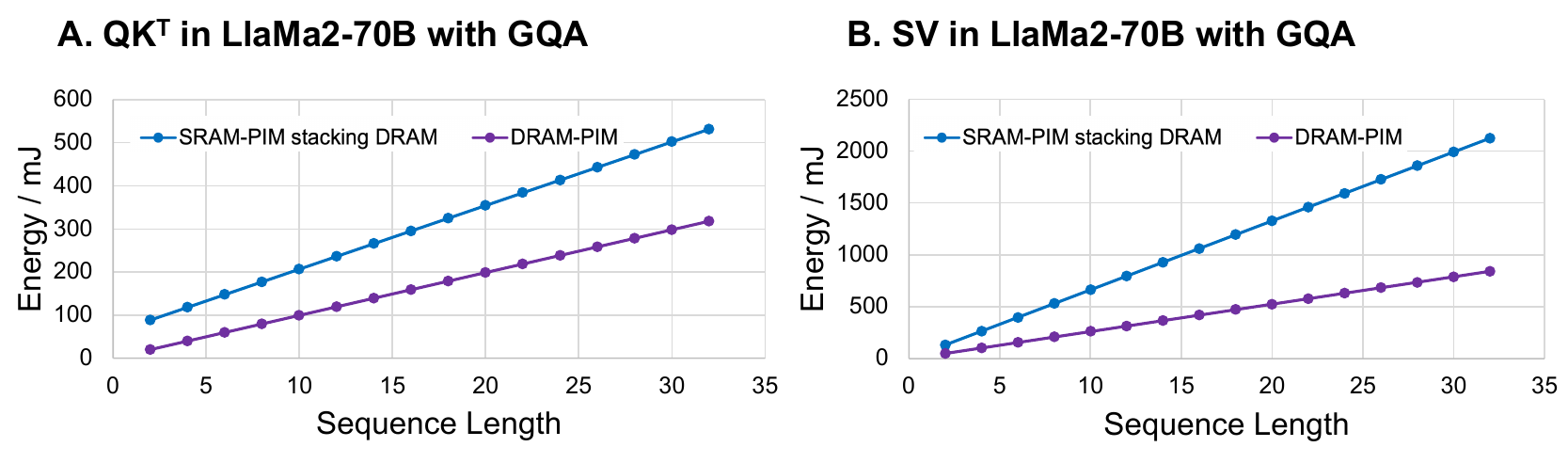}
  \caption{\textcolor{black}{The energy variation between SRAM-PIM stacking DRAM and pure DRAM-PIM.}}
  \label{fig:gqa_energy}
\end{figure}

\textcolor{black}{In $QK^T$, longer sequence and the fewer TPs lead to better reusing of SRAM-PIM. In $SV$, the longer sequence brings more weight reloading, the advantage of SRAM-PIM is limited. However, Fig. \ref{fig:gqa_energy} demonstrates that longer sequence length inevitably results in more cross-die data transfers and higher energy when using SRAM-PIM. For GQA, whether $QK^T$ uses SRAM-PIM for better performance depends on the specific parallelism strategy and sequence length, but for $SV$ DRAM-PIM still has a significant energy advantage.}


\textcolor{black}{Furthermore, CompAir’s value lies not only in demonstrating that PIM systems can achieve competitive energy-efficiency and performance for LLM, but also in proposing a scalable data-centric system. CompAir actually takes LLM as an example, deconstructs different PIM technology paths into vector and matrix-friendly, and adds the CompAir-NoC to realize fine-grained scalar operations, thus constructing a blueprint for energy-efficient computation of vectors, matrices, and scalars at the right place. In this design framework, NVM-PIM replacing SRAM-PIM with adapting better configuration, investigating the mapping between DRAM-PIM and SRAM-PIM are topics worth digging deeper into.}

%% file: Sections/related.tex
\section{Related Works}
\label{sec:bgd_pim}





Recently, commercial DRAM-PIM systems emerge, including Samsung FIMDRAM\cite{FIMDRAM}, UPMEM\cite{UPMEM}, and SK-Hynix AiM\cite{AiM} system. DRAM-PIM can perform massive parallel computing using SIMD vector operations up to 32KB\cite{MIMDRAM}, latest architectures leverage DRAM-PIM for memory-bound tasks in the LLM\cite{NeuPIMs, gu2025cent, lolpim, IANUS, samsung-isscc25-keynote}. To further extend the bandwidth, \cite{TETRIS, Neurocube, AttAcc, ISSCC21-25-4, AttAcc} implement multi-layer DRAM banks vertically via 3D Memory, but at the cost of significantly high power consumption. However, massive SIMD parallelism raises flexibility overhead and the performance of DRAM-PIM heavily rely on suitable mapping and programming\cite{PIMFlow, Cinnamon}, the mismatching causes performance degradation due to inter-bank communication and layout rearrangement\cite{livia, NDPBridge, UMPIM}.

The SRAM-PIM, by integrating the compute logic in/near the SRAM array, enables matrix computation with low-latency in 10ns and 100 TFLOPS/W power efficiency\cite{ISSCC25-14-5, ISSCC25-14-3}. However, the size of a single macro of SRAM-PIM is limited\cite{PIM4AI} and the performance advantage is depend on efficient weight reuse. For LLM, SRAM-PIM needs to scale out\cite{HBA, AIG-CIM}, posing a programmability challenge for MIMD accelerators like Simba\cite{simba}. Moreover, the matrix in attention varies in each inference, SRAM-PIM suffers from frequent swap-out and can hardly achieve a good performance. In all, DRAM-PIM and SRAM-PIM are all promising technologies, but with different advantages. In this work, we try to be compatible with the advantages of both.

\textcolor{black}{In-transit computing has been pioneered in general-purpose processors with two goals: (i) offloading CPU workloads\cite{Sangaiah2020SnackNoC}, (ii) reducing the data movement\cite{Huang2019ActiveRouting, OmniComp}. A similar idea has emerged in memory system, with the objective of performing computation while data are moved across memory hierarchies\cite{livia, tako}, thus avoiding the need for all data to be frequently shuttled between DRAM and CPU pipelines. CompAir-NoC draws on the ideas with novel microarchitecture design, as the first attempt for LLM and PIMs.}